\documentclass[twocolumn,english,aps,superscriptaddress,prb]{revtex4-1}
\usepackage{color}
\usepackage{amsmath}
\usepackage{graphicx}
\usepackage{amssymb}
\usepackage{bm}
\newcommand{\nio}{Na$_2$IrO$_3$}

\newcommand{\rucl}{$\alpha$-RuCl$_3$}

\newcommand{\w}{\omega}
\newcommand{\e}{\epsilon}

\newcommand{\KK}{K}   % Kitaev coupling
\newcommand{\JK}{J_{\rm K}}   % Kondo coupling
\newcommand{\TK}{T_{\rm K}}   % Kondo temperature
\newcommand{\nc}{n_c}   % Occupation number

\newcommand{\ket}[1]{\left| #1 \right\rangle}

\newcommand{\Ztwo}{\mathbb{Z}_2}

\newcommand{\Uone}{\mathrm{U(1)}}
\newcommand{\SUtwo}{\mathrm{SU(2)}}
\newcommand{\SOfour}{\mathrm{SO(4)}}

\newcommand{\HH}{\mathcal{H}}
\newcommand{\Ht}{\mathcal{H}_t}
\newcommand{\HK}{\mathcal{H}_K}
\newcommand{\HJ}{\mathcal{H}_J}

\newcommand{\Xp}{\ket{X_+}}
\newcommand{\Xm}{\ket{X_-}}
\newcommand{\Yp}{\ket{Y_+}}
\newcommand{\Ym}{\ket{Y_-}}
\newcommand{\Zp}{\ket{Z_+}}
\newcommand{\Zm}{\ket{Z_-}}

\newcommand{\Xpm}{\ket{X_\pm}}
\newcommand{\Ypm}{\ket{Y_\pm}}
\newcommand{\Zpm}{\ket{Z_\pm}}

\newcommand{\Xmp}{\ket{X_\mp}}
\newcommand{\Ymp}{\ket{Y_\mp}}
\newcommand{\Zmp}{\ket{Z_\mp}}

\newcommand{\Tone}{\mathrm{T}_1}
\newcommand{\Ttwo}{\mathrm{T}_2}

\newcommand{\eu}{\mathrm{e}}
\newcommand{\iu}{\mathrm{i}}
\let\arrvec=\vec
\renewcommand{\vec}[1]{\bm{#1}}
\newcommand{\mat}[1]{\bm{#1}}
\DeclareMathOperator{\tr}{tr}

\graphicspath{{./}{./figs/}}

\begin{document}

\title{
Fractionalized Fermi liquids and exotic superconductivity in the Kitaev-Kondo lattice
}

\author{Urban F.~P. Seifert}
\author{Tobias Meng}
\affiliation{Institut f\"ur Theoretische Physik, Technische Universit\"at Dresden,
01062 Dresden, Germany}
\author{Matthias Vojta}
\affiliation{Institut f\"ur Theoretische Physik, Technische Universit\"at Dresden,
01062 Dresden, Germany}
\affiliation{Center for Transport and Devices of Emergent Materials, Technische Universit\"at Dresden,
01062 Dresden, Germany}

%%%%%%%%%%%%%%%%%%%%%%%%%%%%%%%%%%%%%%%%%%%%%%%%%%%%%%%%%%%%%%%%%%%%%%%

\begin{abstract}
Fractionalized Fermi liquids (FL$^\ast$) have been introduced as non-Fermi-liquid metallic phases, characterized by coexisting electron-like charge carriers and local moments which itself form a fractionalized spin liquid. Here we investigate a Kondo lattice model on the honeycomb lattice with Kitaev interactions among the local moments, a concrete model hosting FL$^\ast$ phases based on Kitaev's $\Ztwo$ spin liquid. We characterize the FL$^\ast$ phases via perturbation theory, and we employ a Majorana-fermion mean-field theory to map out the full phase diagram. Most remarkably we find nematic triplet superconducting phases which mask the quantum phase transition between fractionalized and conventional Fermi liquid phases. Their pairing structure is inherited from the Kitaev spin liquid, i.e., superconductivity is driven by Majorana glue.
\end{abstract}

\date{\today}


\maketitle
%%%%%%%%%%%%%%%%%%%%%%%%%%%%%%%%%%%%%%%%%%%%%%%%%%%%%%%%%%%%%%%%%%%%%%%

\section{Introduction}

Metals with strong electronic correlations can host a variety of fascinating phases, including unconventional spin and charge density waves as well as high-temperature superconductivity. In addition, they often show marked deviations from the Fermi-liquid phenomenology. These deviations can have various sources: anomalously low coherence temperatures, nearby quantum critical points, quenched disorder, or they can be the property of genuine non-Fermi-liquid phases.\cite{schofield99,stewart01,hvl,ss_strange}

While stable non-Fermi-liquid behavior is generic to one-dimensional interacting electrons, theoretically well-established examples in higher dimensions are rare. One is given by fractionalized Fermi-liquid phases, dubbed FL$^\ast$. Motivated by heavy-fermion non-Fermi liquids, FL$^\ast$ were originally proposed as phases of Kondo-lattice models where Kondo screening is ineffective and the local moments form a fractionalized spin-liquid state instead.\cite{flst1,flst2} In the context of multiorbital or multiband Hubbard models, an FL$^\ast$ phase is an orbital-selective Mott phase where a subset of bands have undergone a Mott transition.\cite{anisimov02,osmott_rev} A defining characteristic of FL$^\ast$ is the presence of a Fermi surface of conventional charge-$e$ spin-$1/2$ quasiparticles which, however, encloses a momentum-space volume determined by conduction electrons alone and therefore, in general, violates Luttinger's theorem in a quantized fashion.

More recently, fractionalized Fermi liquids have been suggested as candidate phases for the pseudogap regime of underdoped cuprates,\cite{qiss10,moonss11,wen11} a concept which was inspired by early ideas of cuprates being described as doped spin liquids.\cite{pwa87} The main difference to the heavy-fermion case is that, for a one-band description of cuprates, local moments and doped holes co-exist in the same band, such that an asymptotic decoupling of the two FL$^\ast$ components cannot be achieved.

\begin{figure}[!tb]
\includegraphics[width=.8\columnwidth,clip]{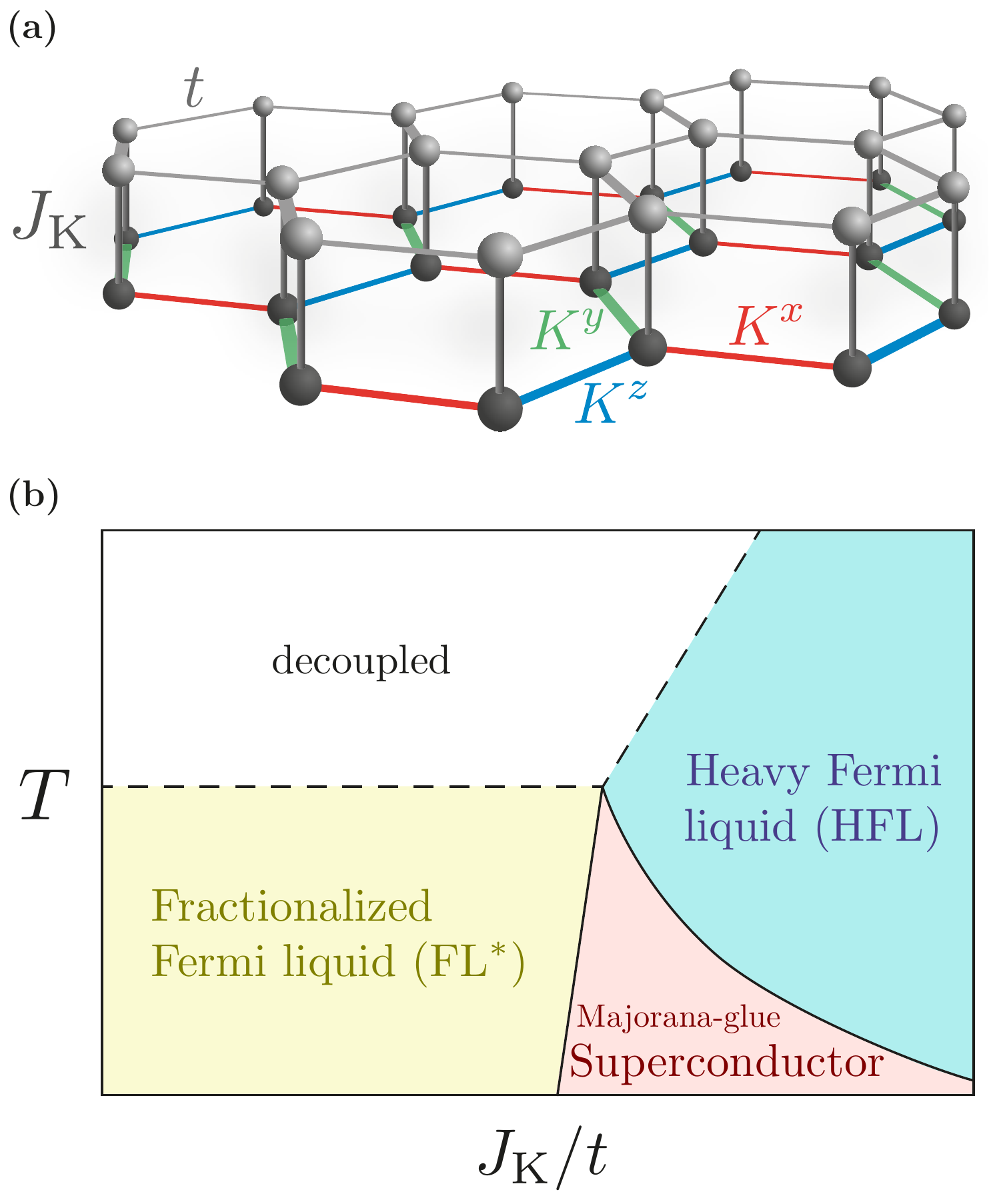}
\caption{
(a) Sketch of the Kitaev-Kondo-lattice model: Conduction electrons move on a honeycomb lattice with hopping energy $t$ (upper layer) and are coupled locally, via a Kondo interaction $\JK$, to spins which interact among themselves via compass interactions $K^{x,y,z}$ (lower layer).
(b) Schematic phase diagram of the Kitaev-Kondo lattice as a function of Kondo coupling $\JK$ and temperature $T$, keeping the conduction-band filling $\nc$ and the Kitaev coupling $\KK$ fixed. Solid lines are symmetry-breaking phase transitions, while dashed lines denotes crossovers (which become phase transitions in the mean-field treatment), for details see text.
}
\label{fig:model_pd}
\end{figure}

While the proof of the existence of fractionalized Fermi liquids in two-band models only relies on the existence of fractionalized spin liquids, concrete calculations have been mainly restricted to toy models, and discussions for more realistic lattices and interactions are scarce. Given that the last decade has seen tremendous progress in finding and characterizing spin-liquid states in concrete microscopic settings,\cite{lee08,balents10} it is a timely issue to close this gap -- this is the purpose of this paper.

%\subsection{Summary of main results}

To this end, we will utilize Kitaev's model for a $\Ztwo$ spin liquid on the honeycomb lattice \cite{kitaev06} and augment this by a band of conduction electrons, with a Kondo-type coupling between the electrons and the local moments, Fig.~\ref{fig:model_pd}(a). The resulting Kitaev-Kondo lattice hosts an FL$^\ast$ phase which can be treated in a controlled fashion, and we discuss its key properties. We then use a Majorana-based mean-field theory to study the full phase diagram. Most interestingly, we find emergent exotic superconducting states at intermediate coupling: These states mask the quantum phase transition between the fractionalized Fermi liquid at small Kondo coupling and a more conventional heavy Fermi liquid at large Kondo coupling, Fig.~\ref{fig:model_pd}(b). The superconducting states display triplet pairing, break discrete rotation and reflection symmetries of the underlying model, and show accidental nodes in the excitation spectrum over significant portions of parameter space. We argue that many of these unconventional properties are inherited from the Kitaev spin liquid, i.e., emerge from those of the matter Majorana fermions of the Kitaev model. This demonstrates that superconductivity here is driven by ``Majorana glue''.

We note that superconducting phases have also been obtained in mean-field\cite{ykv12,hya12,oka13,liu16} and approximate RG\cite{scher14} studies of the doped Heisenberg-Kitaev model. However, the character of those superconducting phases is significantly different from the ones of the present work, as will become clear in the course of the paper.

%\subsection{Outline}

The remainder of the paper is organized as follows:
In Section~\ref{sec:model} we introduce the Kitaev-Kondo lattice model.
In Section~\ref{sec:pert} we discuss the physics of the FL$^\ast$ phase using perturbation theory in the Kondo coupling.
Section~\ref{sec:mf} describes the Majorana-based mean-field theory and discusses aspects of gauge redundancies and projective symmetries.
Section~\ref{sec:mfres} is devoted to the results of the mean-field treatment, i.e., the mean-field phases and phase diagrams. In particular, we will highlight the properties of the emergent superconducting phases.
A discussion and outlook will close the paper.
Technical details, including a full symmetry analysis of the model, are relegated to the appendices.

%%%%%%%%%%%%%%%%%%%%%%%%%%%%%%%%%%%%%%%%%%%%%%%%%%%%%%%%%%%%%%%%%%%%%%%

\section{Kitaev-Kondo-lattice model}
\label{sec:model}

For definiteness, we consider a Kondo-lattice model where both conduction and local-moment electrons live on a two-dimensional bipartite honeycomb lattice, see Fig.~\ref{fig:model_pd}(a). The key ingredient is the compass (or Kitaev) interaction between the local moments.\cite{kitaev06} The total Hamiltonian is $\HH=\Ht+\HK+\HJ$ with
\begin{align}
\label{h}
\Ht &= - t \sum_{\langle ij\rangle\sigma} (c_{i\sigma}^\dagger c_{j\sigma} + h.c.), \notag \\
\HK & = -  \sum_{\langle ij\rangle_\alpha} \KK^\alpha S_i^\alpha S_j^\alpha,  \notag \\
\HJ & =  \frac{1}{2} \sum_{i\sigma\sigma'\alpha} \JK^\alpha c_{i\sigma}^\dagger \tau^\alpha_{\sigma\sigma'} c_{i\sigma'} S_i^\alpha
\end{align}
in standard notation.
The first term represents the conduction-electron kinetic energy, the second the Kitaev coupling among the spin-$1/2$ local moments, with $\langle ij\rangle_\alpha$ denoting an $\alpha$ bond on the lattice ($\alpha=x,y,z$), and the last term represents the local Kondo coupling, with $\tau^\alpha$ the vector of Pauli matrices.
A chemical potential $\mu$ is applied to the conduction electrons to control their filling,
\begin{equation}
\label{filling}
\nc = \frac{1}{N}\sum_{i\sigma} c_{i\sigma}^\dagger c_{i\sigma}\,,
\end{equation}
where $N$ is the number of unit cells. We note that $\nc=2$ corresponds to the ``half-filled'' case where, in the absence of a coupling to the local moments, the chemical potential is at the Dirac point. We will concentrate on $\nc \geq 2$; phases for $\nc \leq 2$ are related by particle-hole symmetry.

The Kitaev model $\HK$ alone describes an exactly solvable $\Ztwo$ spin liquid.\cite{kitaev06} Its degrees of freedom are itinerant ``matter'' Majorana fermions and static $\Ztwo$ gauge fluxes. The matter-Majorana spectrum is gapless and of Dirac type for isotropic couplings, $\KK^x=\KK^y=\KK^z\equiv\KK$, but acquires a gap for large anisotropies. In this paper, we will assume isotropic Kitaev couplings as well as isotropic Kondo couplings, $\JK^x=\JK^y=\JK^z\equiv\JK$, unless otherwise noted.

In analogy to earlier work, \cite{flst1} the Kitaev-Kondo-lattice model $\HH$ \eqref{h} is expected to host a fractionalized Fermi liquid for $\JK\ll\KK,t$ because the Kitaev spin liquid is stable against a small coupling to conduction electrons. Conversely, the model is expected to realize a heavy Fermi liquid for $\KK \ll \JK \sim t$ (or $\KK \ll \TK$ where $\TK$ is the Kondo temperature) dueKitaev-Kondo to robust Kondo screening of the local moments.

The symmetry properties of the model \eqref{h}, with isotropic Kitaev and Kondo couplings, are dictated by the symmetries of the Kitaev model $\HK$. Its spin structure breaks continuous $\SUtwo$ spin rotation symmetry, but combinations of spin and lattice transformations are discrete symmetries of the model.\cite{ykv12,rrtrp15}
A full analysis, presented in Appendix \ref{sec:sym_appendix}, shows that the symmetries at the $K$ point generate the symmetric group $\mathcal S_4$.

%%%%%%%%%%%%%%%%%%%%%%%%%%%%%%%%%%%%%%%%%%%%%%%%%%%%%%%%%%%%%%%%%%%%%%%

\section{Fractionalized Fermi liquids at small Kondo coupling}
\label{sec:pert}

We start the analysis of the Kitaev-Kondo-lattice model \eqref{h} by considering the limit of small Kondo coupling $\JK$. For $\JK=0$ we have two non-interacting subsystems described by $\Ht$ and $\HK$ alone. Perturbation theory in $\JK$ is regular, as the $\Ztwo$ spin liquid described by $\HK$ is protected by its gap to $\Ztwo$ flux excitations, hence small $\JK$ is an irrelevant coupling. The resulting phase is a fractionalized Fermi liquid, and we analyze its properties perturbatively.

\subsection{Effect of $\JK$ on spin liquid}

First, we discuss how the Kondo coupling modifies the properties of the spin-liquid component. The perturbation theory is organized in powers of $\HJ$ and hence in the number of electron--spin interactions: The connected diagrams at $n^{\rm th}$ order in perturbation theory represent processes in which an electron interacts $n$ times with the local moments.
The Hilbert space of $\HK$ alone can be divided into flux sectors which are separated by energy gaps of order $\KK$. Focussing at low energies, we restrict our attention to the lowest (flux-free) sector by projecting the effect of the perturbation back into the flux-free state, i.e., states with excited fluxes may only occur as virtual intermediate states.\cite{kitaev06}

Inspecting the connected diagrams at any order in perturbation theory, we find that the Kondo coupling induces retarded spin exchanges between the local moments that the electron has interacted with in the respective process. The form of these exchange couplings is strongly restricted by the requirement to return the system to the flux-free sector. Because the interaction $\sum_{\sigma\sigma'}\JK^\alpha c_{i\sigma}^\dagger \tau^\alpha_{\sigma\sigma'} c_{i\sigma'} S_i^\alpha$ creates two fluxes in the hexagons next to the $\alpha$-bond of site $i$, it is for example clear that there is no process of first order in $\JK$ that keeps the system in the flux-free sector. Since, in addition, the system has time-reversal symmetry $\mathcal{T}$, which flips the spins, $\mathcal{T}S_i^\alpha\mathcal{T}^{-1}=-S_i^\alpha$, we more generally find that the instantaneous part of all exchange couplings involving an odd number of spins must vanish.

The first non-vanishing contribution in perturbation theory is thus second order in $J_{\rm K}$. To leave the system in the flux-free sector, the second electron-spin-interaction needs to annihilate the fluxes created by the first one. We find that the Kondo coupling then simply renormalizes the Kitaev couplings by a correction of the order of $\JK^2/{\rm max}(t,K)$, see Appendix \ref{append:perturbation_theory}. In higher orders, the Kondo coupling leads to couplings involving a larger number of matter Majoranas. Besides processes that correspond to the creation and subsequent annihilation of pairs of fluxes at different locations in the lattice, there are also ring exchange couplings (at sixth order in perturbation theory, there is for example a processes involving the six local moments around a hexagon), and processes in which fluxes are subsequently created and annihilated at all hexagons alongside paths through the lattice that induce long-range hoppings for the matter Majoranas.

We conclude that the lowest flux sector will be described by weakly interacting matter Majorana fermions with renormalized dispersion. Importantly, for isotropic Kitaev couplings the Majorana spectrum will remain gapless for finite small $\JK$ at any order in perturbation theory: The Dirac points are protected by a combination of time-reversal and lattice symmetries.

Beyond the flux-free sector, the Kondo coupling leads to dynamics for the fluxes (visons): spin flips between electrons and local moments allow the fluxes to hop. For a time-reversal symmetric system, the Kondo coupling induces a (gapped) vison dispersion at order $\JK^2$.

In addition, the conduction electron mediate a Ruderman-Kittel-Kasuya-Yosida (RKKY) interaction between the local moments which -- by analogy to graphene -- scales as $\JK^2$ and decays as $1/R^3$ ($1/R^2$) for $\nc=2$ ($\nc\neq2$),\cite{bs_10,ss_11,ss_11_2} where $R$ is the distance between two local moments. This interaction implies that spin correlations become generically long-ranged, $\langle \arrvec S_i\cdot\arrvec S_j\rangle \propto 1/R_{ij}^3$ ($1/R_{ij}^2$) for $\nc=2$ ($\nc\neq2$).

Importantly, the generated interactions will not destabilize the underlying spin liquid: Spontaneous symmetry breaking is suppressed for small $\JK$ because of the vanishing Majorana density of states in the low-energy limit. This follows from the analogy to graphene, which remains a gapless semimetal even in the presence of long-range Coulomb interactions.\cite{graphene_review}

\subsection{Effect of $\JK$ on conduction electrons}

Second, we discuss the scattering of conduction electrons off local moments, restricting our attention to small $\JK$ and low $T$. Instead of using bare perturbation theory, we account for higher-order effects by noting that the local-moment operator in general acquires a decay channel into two matter Majoranas.\cite{syb16} This yields the most important low-energy scattering process for $c$ electrons, with a self-energy ${\rm Im} \Sigma_c$ scaling as $\w^4$ for the gapless Kitaev model because of the Dirac nature of the matter Majoranas; this is subleading compared to interaction effects among the $c$ electrons. Trivially, for the anisotropic gapped Kitaev model, low-energy scattering is fully absent.

\subsection{Thermodynamic and transport properties}

Finally, the Kondo coupling also constitutes a subleading perturbation for the low-temperature thermodynamics such as specific heat, simply because the density of states of the matter Majoranas vanishes linearly at their Dirac node (again for the gapless Kitaev model), while the conduction electrons have a finite density of states at the Fermi level (we assume a filling $\nc\neq 2$ of the conduction electrons, i.e. away from half-filling, in the remainder). Similarly, the Wiedemann-Franz law should hold: even in the presence of weak disorder, which induces a finite density of states, we expect the thermal conductivity of the matter Majoranas to go to a universal constant,\cite{lee93} while the thermal conductivity of the metallic conduction electrons diverges for divergent scattering times.

\subsection{How topological is a fractionalized Fermi liquid?}

Given that an FL$^\ast$ phase is based on a fractionalized spin liquid, it is worth asking which of its topological properties it inherits. To keep the following discussion simple, we concentrate on an FL$^\ast$ phase derived from a {\em gapped} $\Ztwo$ spin liquid, i.e., having in mind the gapped anisotropic Kitaev model, but most of the following applies more generally.

(i) Any FL$^\ast$ phase displays a Fermi surface whose momentum-space volume is given by that of the conduction electrons alone,
\begin{equation}
{\cal V}_{\rm FL^\ast} = K_d (n_c\,{\rm mod}\,2)
\end{equation}
where $K_d = (2\pi)^d/(2 v_0)$ is a phase space factor, with $v_0$ the unit cell volume, and the factor of 2 accounts for the spin degeneracy of the bands.
In contrast, in a Fermi liquid the Fermi volume is determined by the total number of electrons,
\begin{equation}
{\cal V}_{\rm FL} = K_d (n_{\rm tot}\,{\rm mod}\,2)
\end{equation}
with $n_{\rm tot}=n_c+n_f$ where $n_f$ is the number of local-moment electrons per unit cell.
Hence, FL$^\ast$ is in general characterized by a quantized violation of Luttinger's theorem. We note that, in the present case of a honeycomb lattice, ${\cal V}_{\rm FL^\ast}={\cal V}_{\rm FL}$ because $n_f=2$.

(ii) FL$^\ast$ is characterized by non-trivial emergent excitations of the spin-liquid component, in addition to conventional electronic quasiparticles. In fact, it is these excitations which enable a violation of Luttinger's theorem.\cite{flst1,oshi00}

(iii) The existence of gapped vison excitations, which protect FL$^\ast$ and which cannot be created individually by any local operator, implies the existence of topologically distinct sectors if placed on a torus. These sectors are distinguished by visons selectively threaded through the torus holes, with degenerate ground states in the thermodynamic limit. Note, however, that the coupling between the sectors does not scale exponentially to zero with increasing system size, as is the case in a fully gapped spin liquid. In FL$^\ast$, the fact that correlation functions become in general long-ranged changes the finite-size scaling of the total energy, i.e., finite-size corrections are generically of power-law type.

(iv) The combination of the spin-liquid and conduction-electron components can be expected to lead to violations of the area law of the entanglement entropy \cite{arealaw}, with details depending on the nature of the underlying spin liquid. A detailed study of this is left for future work.

%%%%%%%%%%%%%%%%%%%%%%%%%%%%%%%%%%%%%%%%%%%%%%%%%%%%%%%%%%%%%%%%%%%%%%%

\section{Majorana-fermion mean-field theory}
\label{sec:mf}

The model $\HH$ \eqref{h} is not exactly solvable in the presence of a finite Kondo coupling. In order to go beyond perturbation theory, we pursue an approximate solution using a suitable mean-field approach. In contrast to most mean-field treatments of Kondo-lattice models in the literature, the Majorana mean-field theory described below has the advantage that it is {\em exact} in the $\JK=0$ case, i.e., it correctly reproduces the physics of the Kitaev spin liquid.

\subsection{Majorana representation} \label{subsec:majrep}

Spin liquids are commonly studied by representing the spin operator of local moments in terms of slave fermions $f_{j \sigma}$ as $S^\alpha_i = f_{i\sigma}^\dagger \tau^\alpha_{\sigma \sigma'} f_{i \sigma} /2$ along with the local single-occupancy $n_{i \uparrow} + n_{i \downarrow} = 1$ constraint.
There is a $\SUtwo$ gauge redundancy in the above description which amounts to taking the Nambu spinor $(f_\uparrow, f_\downarrow^\dagger)^T \mapsto W (f_\uparrow, f_\downarrow^\dagger)^T$ for some $W \in \SUtwo$.\cite{wen08}

The Kitaev model $\HK$, however, can be solved exactly by using a representation of a local moment $S^\alpha_i = \iu \chi^0_i \chi^\alpha_i$ in terms of Majorana fermions $\chi^\mu_i = {\chi^\mu_i}^\dagger$ (in real space and $\mu = 0, \dots, 3$) with the anticommutation relations $\{\chi^\mu_i, \chi^\nu_j\} = \delta^{\mu \nu} \delta_{i j}$,%
\cite{fn:maj_norm} %
 and the local constraint $D_i = 4 \chi^0_i \chi^1_i \chi^2_i \chi^3_i = 1$ for physical states.\cite{kitaev06}

It has been pointed out\cite{ykv12} that by decomposing the slave fermions into Majorana fermions one can obtain Kitaev's representation of the spin operators. Specifically, once can choose $f_\uparrow = (\chi^0 + \iu \chi^3) / \sqrt{2}$ and $f_\downarrow = (\iu \chi^1 - \chi^2) / \sqrt{2}$ and obtain\cite{fn:mapping}
\begin{equation} \label{eq:spin_maj}
  S^\alpha_i = \frac{\iu}{4} \left( \chi^0_i \chi^\alpha_i - \chi^\alpha_i \chi^0_i - \e^{\alpha \beta \gamma} \chi_i^\beta \chi_i^\gamma \right);
\end{equation}
this is the representation to be used below.
The single-occupancy constraint
\begin{equation}
  0 = n_{i\uparrow} + n_{i\downarrow} - 1 = \iu \chi^0_i \chi^3_i + \iu \chi^1_i \chi^2_i,
\end{equation}
can be shown to generate the operator $D = 4 \chi^0 \chi^1 \chi^2 \chi^3$ with the constraint $D=1$ for physical states (cf. Appendix \ref{sec:constraints}).
For states which fulfill the constraint, the form of $S^\alpha$ given above reduces to the representation used by Kitaev, $S^\alpha = \iu \chi^0 \chi^\alpha$.
Eq.~\eqref{eq:spin_maj} can be written in a more compact manner by introducing the four-vector $\vec \chi$ with components $\chi^\mu$ and the spin operator as $S^\alpha_i = (\iu / 4) \vec{\chi}^T \mat M^\alpha \vec \chi$, where the matrices $\mat M^\alpha \in \SOfour$ are given by
\begin{equation*}
  \mat M^1 = \tau^3 \otimes \iu \tau^2, \quad \mat M^2 = \iu \tau^2 \otimes \tau^0 \ \text{and} \ \mat M^3 = \tau^1 \otimes \iu \tau^2.
\end{equation*}
Note that $[\mat{M}^\alpha,\mat{M}^\beta] =2  \epsilon^{\alpha \beta \gamma} \mat M^\gamma$, so that the matrices $\mat M^\alpha$ furnish a representation of $\SUtwo$. In the following, we will refer to theses matrices as \emph{spin matrices}. Spin rotations can be implemented by transforming $\vec \chi \mapsto \mat R \vec \chi$, where $\mat R \in \SOfour$ is formed by an appropriate linear combination
\begin{equation} \label{eq:general_rot_matrix}
  \mat R_S = a^0 \mat 1 + a^\alpha \mat M^\alpha
\end{equation}
with $a_0^2+ a^\alpha a^\alpha = 1$.

The above mentioned $\SUtwo$ redundancy manifests itself in the present formalism as the invariance of $S^\alpha$ under $\vec \chi \mapsto \mat G \vec \chi$, where $\mat G$ is an $\SOfour$ matrix in the subspace that commutes with $\mat M^\alpha$. A basis for this subspace is given by the matrices
\begin{equation*}
  \mat G^1 = -\tau^0 \otimes \iu \tau^2, \quad \mat G^2 =  -\iu \tau^2 \otimes \tau^3 \ \text{and} \ \mat G^3 = -\iu \tau^2 \otimes \tau^1,
\end{equation*}
where $[\mat G^\alpha, \mat G^\beta] = 2 \epsilon^{\alpha \beta \gamma} \mat{G}^\gamma$, so that these matrices furnish another $\SUtwo$ representation.
Indeed, $\SOfour \simeq \SUtwo \otimes \SUtwo / \Ztwo$. We will refer to the $\mat G^\alpha$ as \emph{isospin matrices}. These matrices can be understood as the Majorana analogue of the Pauli matrices for the Nambu spinor introduced above.

The isospin matrices naturally define the isospin $J^\alpha = (\iu / 4) \vec \chi^T \mat G^\alpha \vec \chi$, which is sometimes also referred to as pseudospin. It is the generator of particle-hole and $\Uone$-charge transformations and first been discussed in the large-$U$ limit of the Hubbard model.\cite{azha88,xusa10}

Note that the constraint amounts to working with isospin singlet states\cite{wen08} with
\begin{equation}
\label{iso_singlet}
\vec \chi^T \mat{G}^\alpha \vec \chi = 0\,.
\end{equation}
Kitaev's representation of the spin operators is obtained in the present formalism by taking
\begin{equation} \label{eq:kitaev_rep}
  S^\alpha = \frac{\iu}{4} \vec \chi^T \left[ \mat M^\alpha - \mat G^\alpha \right] \chi = \iu \chi^0 \chi^\alpha,
\end{equation}
which amounts to including the constraint in each spin operator.\cite{fn:spinmap}
As opposed to the $\mat M^\alpha$ and $\mat G^\alpha$, the matrices $[\mat M^\alpha - \mat G^\alpha]$ do not form a Lie algebra itself, leading to the projective realization of spin rotations (see also Section~\ref{subsec:proj_symm}).

\subsection{Mean-field theory for the Kitaev model}\label{subsec:kitmft}

We start with a mean-field analysis of the pure Kitaev model $\HK$, targeting at paramagnetic solutions. Using the Majorana representation \eqref{eq:spin_maj} and performing a mean-field decoupling, the Hamiltonian now reads
\begin{align}
  \HK &= - K \sum_{\langle i j \rangle_\alpha}  \frac{\iu^2}{4^2} \vec \chi_i^T \mat M^\alpha \vec \chi_i \vec \chi_j^T \mat M^\alpha \vec \chi_j + \sum_{i,\alpha} \lambda^\alpha \iu \vec \chi_i^T \mat G^\alpha \vec \chi_i \notag \\
  &= \frac{\KK}{4} \sum_{\langle ij\rangle_\alpha} \Big[ - \iu \vec \chi^T_i \mat{M}^\alpha \mat{U}_{ij} \mat{M}^\alpha \vec\chi_j  \notag \\
  &~~+ \frac{1}{2} \tr \mat M^\alpha \mat U_{ij} \mat M^\alpha \mat U^T_{ij} \Big] + \sum_{i,\alpha} \lambda^\alpha \iu \vec \chi_i^T \mat G^\alpha \vec \chi_i,
\label{eq:kitaev_mft}
\end{align}
where the real mean fields
\begin{equation}
\label{mfeq1}
U^{\mu \nu}_{ij} = \langle \iu \chi_i^\mu \chi_j^\nu \rangle
\end{equation}
are to be determined self-consistently. We emphasize that instead of using the representation \eqref{eq:kitaev_rep} for the spin operators, we use the more general (gauge-equivalent) expression \eqref{eq:spin_maj} and include Lagrange multipliers $\lambda^\alpha$ to enforce the isospin-singlet constraint on each site.
This allows us to later address mean-field regimes that are inherently different from the Kitaev model (in particular involving the delocalization of the $\chi^1,\chi^2,\chi^3$ Majoranas), as occurring in the full model $\HH$.

Guided by the exact solution of the Kitaev model and previous mean-field treatments,\cite{ykv12} we parametrize the mean-field ansatz for the Kitaev model as $U^{00} = \langle \iu \chi^0_i \chi^0_j \rangle =: u^0$, $U^{\alpha \alpha} = \langle \iu \chi_i^\alpha \chi_j^\alpha \rangle =: u^a$ on $\langle i j \rangle_\alpha$ links, and $U^{\beta \beta} = \langle \iu \chi_i^\beta \chi_j^\beta \rangle =: u^b$ for $\beta \neq \alpha$, yielding
\begin{equation} \label{eq:h_mft_kitaev}
  \HK = \frac{\KK}{4} \sum_{\substack{\langle i j \rangle_\alpha\\ \beta \neq \alpha}} \iu u^a \chi^0_i \chi^0_j + \iu u^0 \chi^\alpha_i \chi^\alpha_j + \iu u^b \chi^\beta_i \chi^\beta_j - u^0 u^a - u^b u^b,
\end{equation}
where we adopt the convention that $i \in A$ sublattice and $j \in B$. Note that we have omitted the Lagrange multipliers since the constraints are automatically satisfied for $\lambda = 0$.
%To compare our results with the exact solution, we hence set in the following $K=4$.

\begin{figure}[!tb]
\includegraphics[width=0.97\columnwidth]{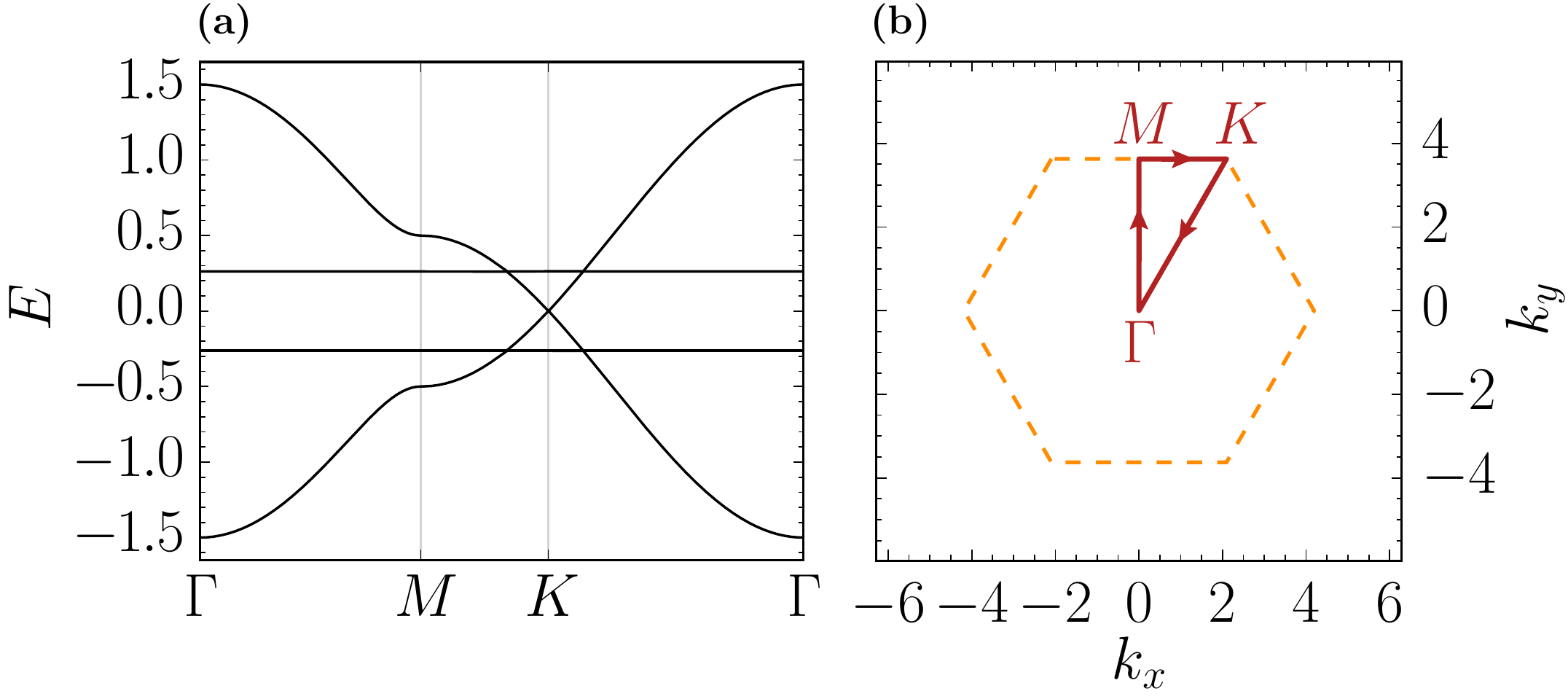}
\caption{
(a): Double spectrum obtained from Majorana mean-field theory for the Kitaev model at $K=4$, see text. The flat bands at $\pm 0.26$ are each threefold degenerate.
(b): First Brillouin zone of the honeycomb lattice, together with the path used in this and subsequent figures.
}
  \label{fig:pureKitaev_bandstruct}
\end{figure}

The resulting Majorana-bilinear Hamiltonian can then straightforwardly be solved in momentum space. Since $\chi_k^\dagger = \chi_{-k}$ for Majorana fermions, the Fourier expansion only extends over half of the Brillouin zone,\cite{cmt94}
\begin{equation}
  \chi_j = \frac{1}{\sqrt{N}}  \sum_{k \in \mathrm{BZ}/2} \left[ \chi_k \eu^{\iu k x_j} + \chi_k^\dagger \eu^{-\iu k x_j} \right],
\end{equation}
so that one obtains a double spectrum with $\varepsilon(k) = - \varepsilon(-k)$ upon diagonalization on one half of the Brillouin zone. By means of a particle--hole transformation one can then recover four Majorana bands on the full Brillouin zone.
This solution can then be used to compute expectation values needed to solve the mean-field equations. Their self-consistent solutions at $T=0$ are given by
\begin{equation}
  u^0(\gamma) = \pm 0.2624, \quad u^a (\gamma) = \mp 0.5 \ \text{and} \ u^b = 0,
\end{equation}
for $\gamma = x,y,z$-bonds, reproducing the mean-field theory found by You et al.\cite{ykv12}
Note that there is $\Ztwo$ redundancy of choosing the signs of $u^0$ and $u^a$ on $\gamma$-bonds as long as $u^0 u^a < 0$. This redundancy can be understood by performing gauge transformations to the gauge field $u_{ij}$ in the exact solution of the Kitaev model.\cite{kitaev06}

Since we chose the spin representation \eqref{eq:spin_maj} [instead of \eqref{eq:kitaev_rep}], the global Kitaev coupling differs by a factor of $1/4$. To compare our results with the exact solution, we henceforth set $K=4$ unless otherwise noted.

The double spectrum for the Kitaev model is shown in Fig.~\ref{fig:pureKitaev_bandstruct}.
The dispersing Majorana mode $\chi^0$ with a graphene-like dispersion
\begin{equation}
  E(k_x,k_y) = \frac{\KK}{4} \left| u^a(x) \eu^{\iu \arrvec k \cdot \arrvec n_1} + u^a(y) \eu^{\iu \arrvec k \cdot \arrvec n_2} + u^a(z) \right|,
\end{equation}
with the lattice vectors of the honeycomb lattice $\arrvec n_{1,2} = (\pm 1, \sqrt{3})^T/2$, is clearly visible.
In addition, we obtain three flat bands associated with the $\chi^1,\chi^2$ and $\chi^3$ Majoranas localized on the respective bonds.

The mean-field theory developed above hence reproduces the exact solution of the Kitaev model. The mean-field parameter $u^a = \langle \iu \chi^\alpha_i \chi^\alpha_j \rangle$ effectively takes the role of the $\Ztwo$ gauge field in the flux-free ground state, as found in the exact solution.
Furthermore, we obtain for the equal-time spin-spin correlators
\begin{equation}
  \langle S_i^\alpha S_j^\alpha \rangle = -\frac{1}{4} \langle \iu \chi^0_i \chi^0_j \rangle \langle \iu \chi^\alpha_i \chi^\alpha_j \rangle \delta_{\langle i j \rangle_\alpha} = -\frac{u^0 u^a}{4} \delta_{\langle i j \rangle_\alpha}.
\end{equation}
This matches the the exact result up to a factor of $1/4$, the latter originating from our choice of the spin representation.

For the most general case with $u^b \neq 0$ (which is of relevance for the further sections), the non-vanishing spin-spin correlation functions for the local moments on neighboring sites can be expressed using the mean-field decoupling as
\begin{align}
    \langle S^\alpha_i S^\alpha_j \rangle &= - \frac{1}{4} \left[u^0 u^a + \left(u^b\right)^2 \right] \quad \text{on $\langle i j \rangle = \alpha$ links, and} \notag\\
      \langle S^\alpha_i S^\alpha_j \rangle &= - \frac{1}{4} \left[u^0 u^b + u^a u^b \right] \quad \text{on $\langle i j \rangle \neq \alpha$ links.}
\label{eq:nn_spincorrel}
\end{align}
Note that the spin correlation functions above are clearly invariant under $\Ztwo$ gauge transformations which flip the sign of the mean fields $u^0,u^a \to -u^0,-u^a$ as detailed above, as long as $u^b = 0$.
A finite value of $u^b$ thus spoils the gauge structure of the Kitaev spin liquid.

For a further discussion of the mean-field theory for the Kitaev spin liquid, in particular for the case of anisotropic couplings, we refer the reader to Appendix \ref{sec:aniso_mft_kitaev}. We note that the Kitaev model can also been treated in a slave-fermion mean-field approximation, as demonstrated by Burnell and Nayak.\cite{buna11} There, the resulting fermion-bilinear mean fields explicitly break the $C_3$ symmetry, requiring a more careful treatment of the gauge transformations needed to obtain a form-invariant Hamiltonian.

\subsection{Mean-field theory for the Kitaev-Kondo lattice}

Since the Kitaev spin liquid is most naturally described via Majorana fermions, it appears to be sensible to introduce a description of the Kitaev-Kondo lattice in terms of Majorana fermions as well.
A Majorana representation of conduction electrons has previously been used in the study of odd-frequency superconductivity.\cite{cmt94} We introduce a phase factor for the $c$ electrons on the $B$ sublattice, $c_B \rightarrow \iu c_B$ and then decompose canonical fermions $c_\uparrow, c_\downarrow$ on each site into four Majorana fermions $\eta^\nu$, using the mapping $c_\uparrow = (\eta^0 + \iu \eta^3) / \sqrt{2}$ and $c_\downarrow = (\iu \eta^1 - \eta^2) / \sqrt{2}$. This amounts to different Majorana representations on the two sublattices, such that the kinetic energy $\Ht$ assumes the simple form (implicit summation over $\lambda = 0, \dots, 3$)\cite{cmt94,bas15}
\begin{align}
    \Ht - \mu \mathcal{N} &= - t \sum_{\langle i j \rangle} \iu \eta_i^\lambda \eta_j^\lambda - \mu \sum_j \left[1 + \iu \left( \eta^0_j \eta^3_j + \eta^1_j \eta^2_j \right) \right] \notag \\
    &= -t \sum_{\langle i j \rangle} \iu \vec \eta_i^T \vec\eta_j - \mu \sum_j \left[ 1 + \frac{\iu}{2} \vec \eta_j^T \mat{G}^3 \vec\eta_j \right] \label{eq:h_maj_kin}
\end{align}
where $\mathcal{N}=\sum_{i\sigma} c_{i\sigma}^\dagger c_{i\sigma}$ is the number of conduction electrons.

Analogous to the decoupling of the quartic Majorana term in Eq.~\eqref{eq:kitaev_mft}, the Kondo interaction $\HJ$ in \eqref{h} in the mean-field approximation assumes the form
\begin{equation} \label{eq:kondo_mft}
  \HJ = \frac{\JK}{4} \sum_{i,\alpha} \left[ \iu \vec\chi_i^T \mat M^\alpha \mat W_i \mat M^\alpha \vec\eta_i - \frac{1}{2} \tr \mat M^\alpha \mat W_i \mat M^\alpha \mat W_i^T \right]
\end{equation}
with the real mean-field parameters
\begin{equation}
\label{mfeq2}
W_i^{\mu \nu} = \langle \iu \chi_i^\mu \eta_i^\nu \rangle
\end{equation}
to be determined self-consistently.

We note that the mean-field decouplings introduced here (and above for the spin liquid) favor paramagnetic solutions with $\langle \arrvec S \rangle=0$ and for the conduction electrons $\langle c^\dagger_\sigma \arrvec \tau_{\sigma \sigma'} c_{\sigma'} \rangle /2 = 0$.

\subsection{Quantum order and projective symmetries}
\label{subsec:proj_symm}

A quantum-ordered spin liquid can be classified in terms of its projective symmetry group (PSG).\cite{wen08} Due to the redundancy in the slave-fermion representation of the spins, physical symmetries (of the projected wavefunction) do not necessarily originate in the symmetry of the ansatz itself, but rather from a transformation that takes the mean-field ansatz to a gauge-equivalent ansatz.

As previously described, spin rotations can act projectively on fermionic partons.\cite{ceh12,ykv12} In the Kitaev model, the spin rotation symmetry is realized by a combination of spin rotation and gauge (isospin) rotation, which has previously been dubbed ``spin-gauge locking''. This is particularly evident from the representation \eqref{eq:kitaev_rep}. To achieve a rotation of the object $\left[\mat M^\alpha - \mat G^\alpha \right]$, a simultaneous spin rotation $\mat{R}_S$ and gauge transformation $\mat R_G$ need to act on the Majorana fermions $\vec \chi$, such that the spin in this representation transforms as
\begin{subequations}
  \begin{align}
    S^\alpha \rightarrow& \frac{\iu}{4} \vec \chi^T \mat R_G^T \mat R_S^T \left[\mat M^\alpha - \mat G^\alpha \right] \mat R_S \mat R_G \vec\chi \\
    &= \frac{i}{4} \vec \chi^T \left[\mat R_S^T \mat M^\alpha \mat R_S - \mat R_G^T \mat G^\alpha \mat R_G  \right] \vec \chi.\label{eq:spin_rot_kitaev}
  \end{align}
\end{subequations}
If the spin rotation matrix $\mat R_S$ is given by $\mat R_S = a_0 \mat{1} + a^\alpha \mat{M}^\alpha$ with $(a^0)^2 + a^\alpha a^\alpha =1$, the isospin transformation matrix $\mat R_G$ required is simply given by
\begin{equation}
  \mat R_G = \pm \left( a^0 \mat{1} + a^\alpha \mat{G}^\alpha \right),
\end{equation}
so that both terms in Eq.~\eqref{eq:spin_rot_kitaev} transform in the same way. Note that there is a residual $\Ztwo$ freedom when choosing the sign of $\mat R_G$, allowing for the classification of the quantum order of the Kitaev model in terms of $\Ztwo$ PSGs.\cite{ykv12}

The conduction band Majorana fermions in Eq.~\eqref{eq:h_maj_kin}, however, do not obey such quantum order. Clearly, the Hamiltonian is invariant under spin rotations $\vec \eta \rightarrow \mat{R}_S \vec \eta$.
In the case of half-filling ($\mu = 0$), we furthermore have the $\SUtwo$ isospin symmetry $\vec \eta \rightarrow \mat{R}_G \vec \eta$ of rotations in the particle-hole and charge sector.
Going away from half-filling, this $\SUtwo$ symmetry is lowered to a residual $\Uone$ symmetry $\vec \eta \rightarrow \mat{R}_C \vec \eta$ with $\mat{R}_C = a^0 \mat{1} + a^3 \mat{G}^3$, which corresponds to particle number/charge conservation.\cite{nils11}

The above considerations allow us to formulate how the decoupling fields $\mat W_i$ in Eq.~\eqref{eq:kondo_mft} transform under a symmetry transformation $\hat{S}$. This transformation may act on the real-space index (such as point group operations), on the local-moment Majoranas $\vec \chi$ and the conduction-band Majoranas $\vec \eta$ as
\begin{equation}
  \hat{S}^{-1} \vec \chi_j \hat{S} = \mat{R}_S^{(\chi)} \mat{R}_G^{(\chi)} \vec \chi_{S(j)}\,,~~~ \hat{S}^{-1} \vec \eta_j \hat{S} = \mat{R}_S^{(\eta)} \vec \eta_{S(j)},
\end{equation}
where we allow for different matrix representations of the transformation for $\vec\chi$ and $\vec\eta$, respectively. Consequently, $\mat W_j$ transforms under the symmetry operation $\hat{S}$ as
\begin{equation}
   \hat{S}^{-1} \mat{W}_j \hat{S} = {\mat{R}_G^{(\chi)}} {\mat{R}_S^{(\chi)}} \mat{W}_{S(j)} {\mat{R}^{(\eta)}_S}^T.
\end{equation}
With the transformation properties of the mean fields $\mat W$ at hand, we will be able to discuss the symmetries of the mean-field phases described in the next section.

%%%%%%%%%%%%%%%%%%%%%%%%%%%%%%%%%%%%%%%%%%%%%%%%%%%%%%%%%%%%%%%%%%%%%%%

\section{Mean-field phases of the Kitaev-Kondo lattice}
\label{sec:mfres}

\begin{figure}[!tb]
\includegraphics[width=\columnwidth,clip]{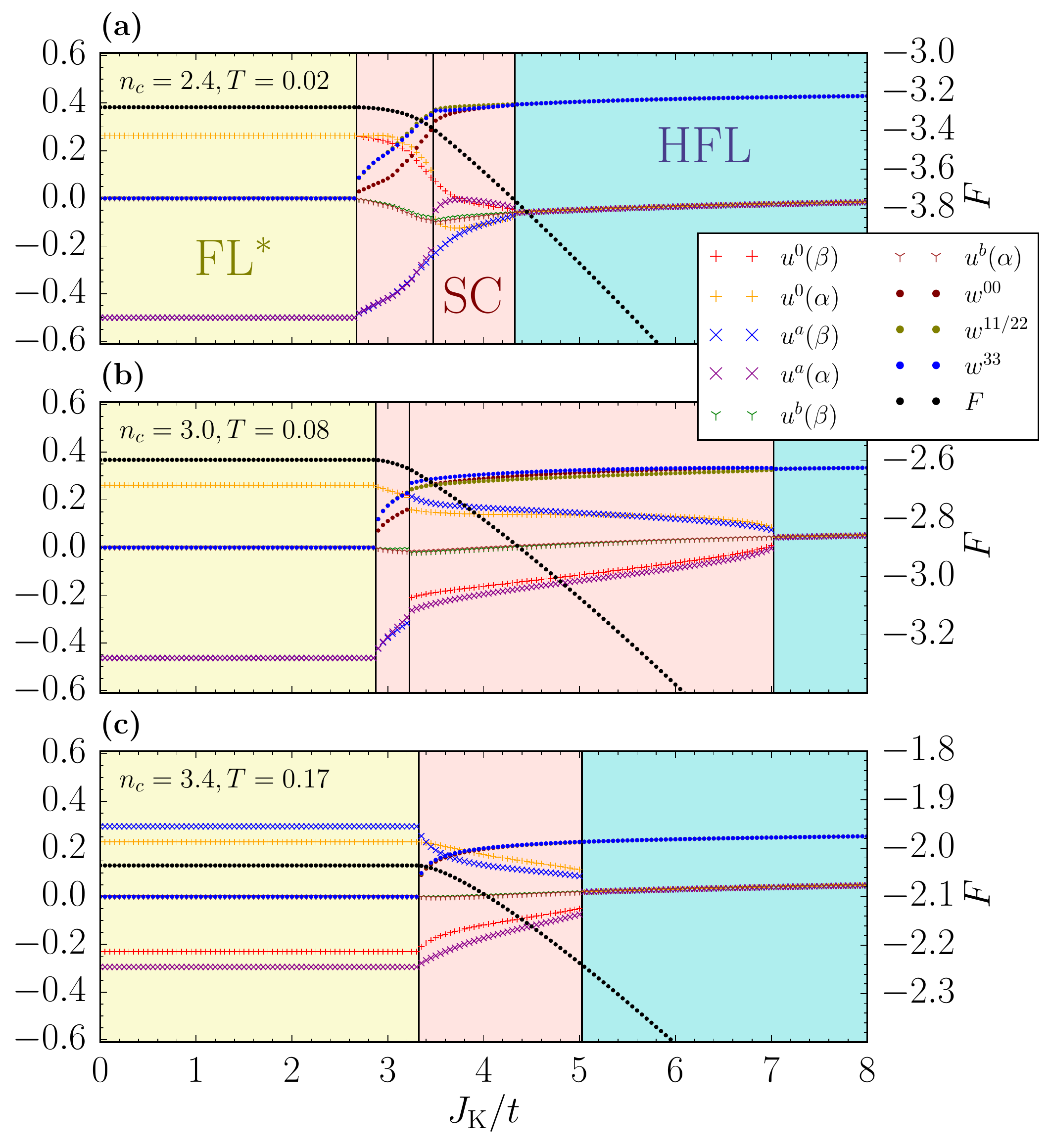}
\caption{Mean-field parameters and Helmholtz free energy $F \equiv U - T S$ as a function of $\JK$ at\cite{fn:zero_temp} $T=0$. For clarity we select a \emph{diagonal} solution (of type $\ket{Z_+}$, see also Section \ref{subsec:sc} and Appendix \ref{sec:majToSF}) in the superconducting phase, thus also fixing a $\Uone$ phase. Note that in this case, always $w^{11} = w^{22}$ holds (cf. discussion succeeding Eq.~\eqref{eq:wz_decomposed}). Here, $\alpha=z$ and $\beta = x,y$.
The conduction-band filling is
(a) $\nc = 2.4$,
(b) $\nc = 3.0$,
(c) $\nc = 3.4$.
In (a) and (b), a first-order transition within the superconducting phase is visible where the nodal structure changes.
In (c) a different gauge for the $u$-parameters has been chosen according to the redundancy described in Section \ref{subsec:kitmft}.
}
\label{fig:mftparams_jk}
\end{figure}

\begin{figure}[!tb]
  \includegraphics[width=\columnwidth,clip]{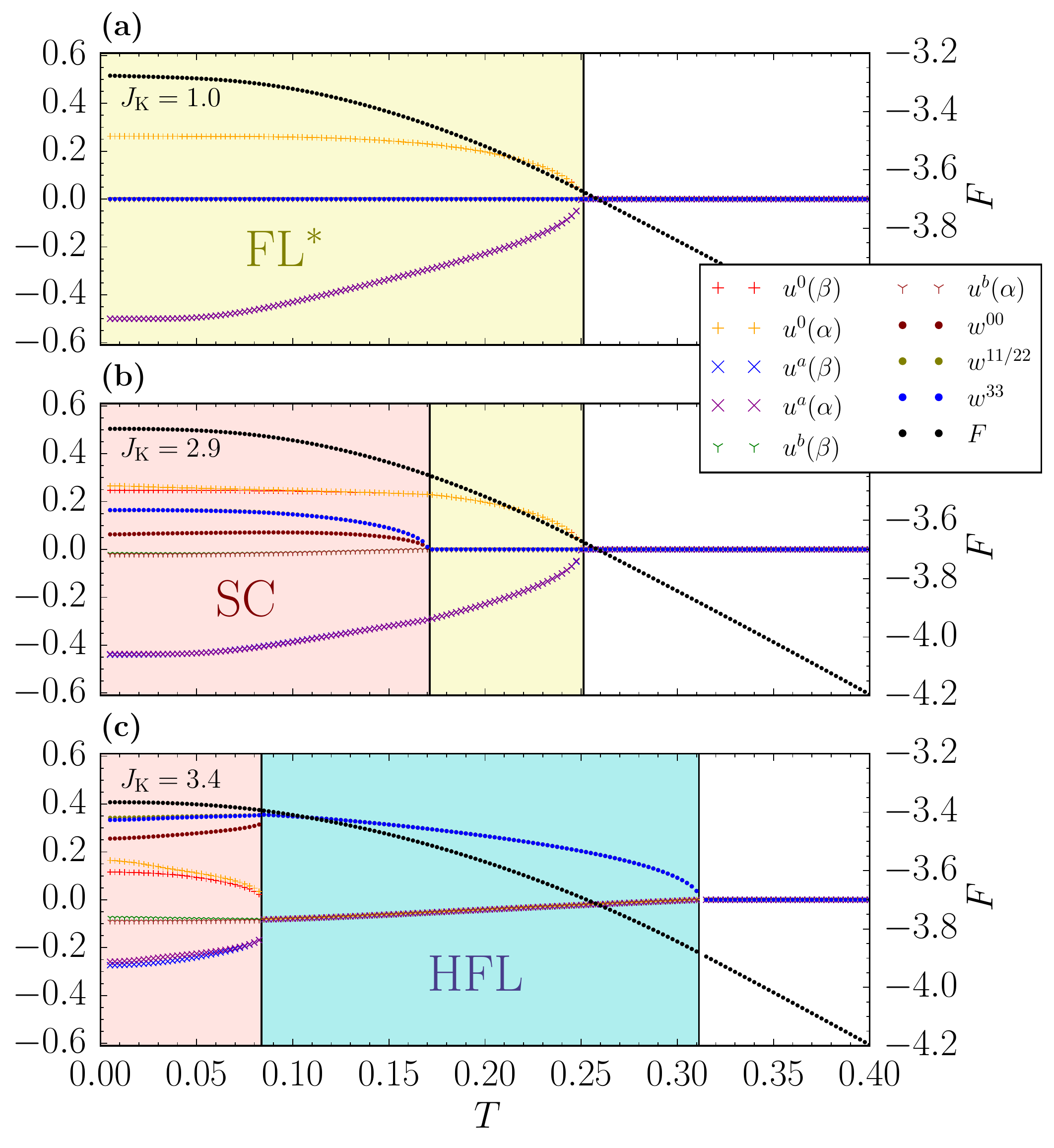}
  \caption{Mean-field parameters as in Fig. \ref{fig:mftparams_jk}, but now as function of temperature $T$ at fixed $n_c=2.4$.
   (a) Transition from FL$^\ast$ to the decoupled regime for $\JK = 1.0$. (b) Transition for $\JK = 2.9$ from SC to HFL (at $T \simeq 0.17$) and then to the decoupled regime (at $T \simeq 0.25$). (c) Transition from HFL to the decoupled regime for $\JK = 3.4$ (at $T\simeq 0.31$).
}
\label{fig:mftparams_t}
\end{figure}

We have solved the mean-field equations \eqref{mfeq1} and \eqref{mfeq2}, together with the constraints \eqref{filling} and \eqref{iso_singlet}, for a range of Kondo couplings $\JK/t$, band fillings $\nc$, and temperatures $T$. We employ units where $t=1$, leading to a $c$-electron bandwidth of $6$, and we set $\KK=4$ unless noted otherwise.
Assuming unbroken lattice translation invariance, the problem involves four chemical potentials and $9+16$ real scalar mean-field parameters ($9$ for $u$ and $16$ for $W$).
To find solutions to the mean-field equations, we have employed an iterative scheme with randomly weighted updates in each step. The iterations were started with different randomly selected initial conditions. If multiple inequivalent solutions occurred, we selected the solution with the lowest (Helmholtz) free energy. Most results have been obtained with a momentum discretization of $16^2$ points; a higher momentum resolution was used to extract spectral properties.

\subsection{Overview}
\label{overview}

A schematic phase diagram is shown in Fig.~\ref{fig:model_pd}(b), and quantitative results are presented in Figs.~\ref{fig:mftparams_jk} and \ref{fig:mftparams_t}. Similarly to earlier works \cite{flst1}, there are four main phases.

(i) At high temperatures there is a decoupled phase, with $\mat{U}=\mat{W}=0$. As discussed in earlier works \cite{flst1}, this decoupling is an artifact of mean-field theory and indicates a regime where conduction electrons are incoherently scattered off local moments.

(ii) At small $\JK/t$ and sufficiently low temperatures, a phase with $\mat U$ mean-field parameters identical to those of the Kitaev spin liquid and vanishing Kondo mean fields, $\mat W = 0$, emerges.
This is the advertised FL$^\ast$ state; for details see Section~\ref{subsec:flst}.

\begin{figure}[!tb]
  \includegraphics[width=\columnwidth,clip]{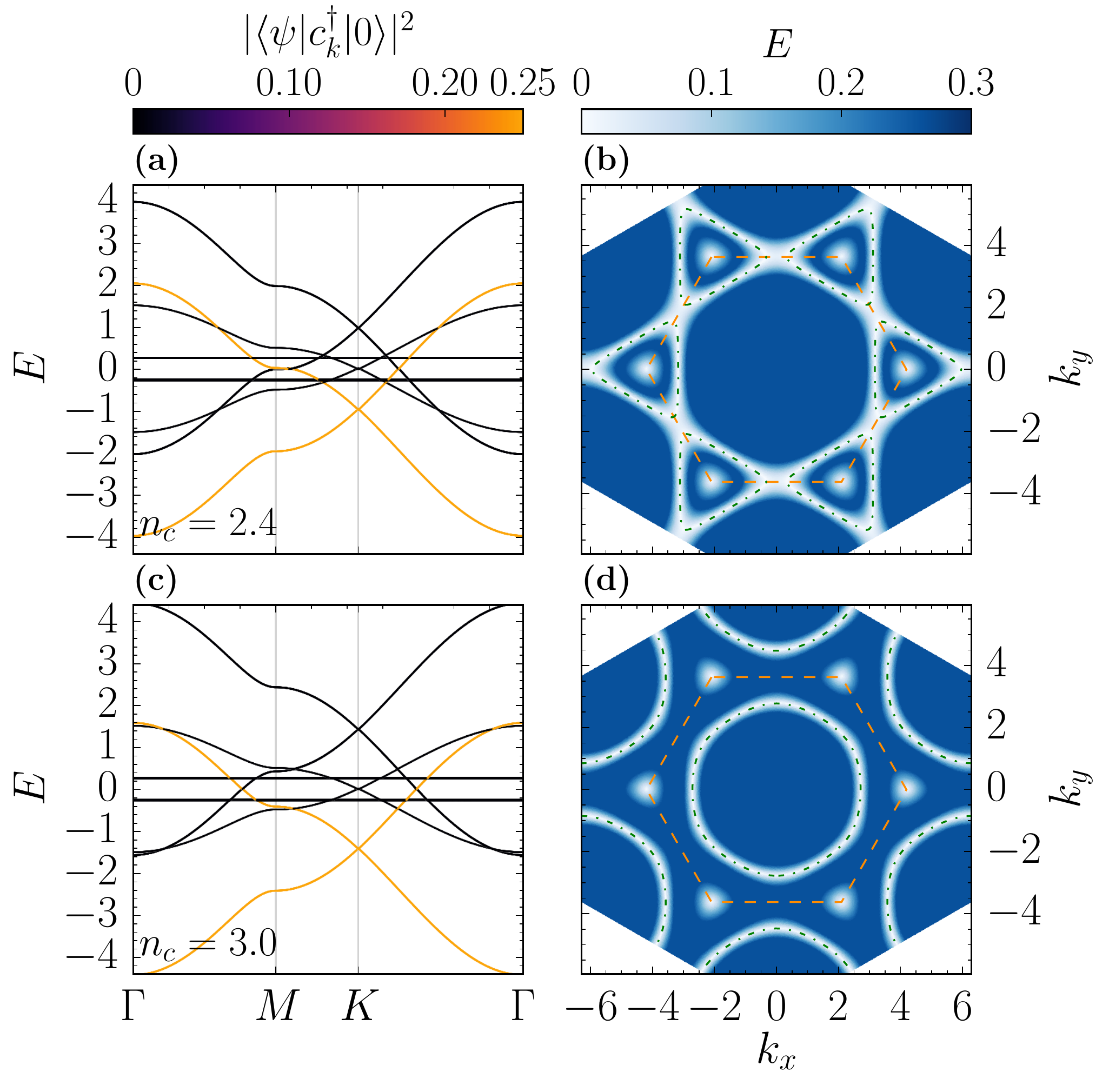}
\caption{Mean-field band structure in the fractionalized Fermi liquid at $\JK = 1.0$.
(a) Cut along high-symmetry lines at $\nc=2.4$ with color-coded quasiparticle weights (averaged over spin and sublattice and normalized to take into account the double spectrum).
(b) Energy of the lowest quasiparticle band at $\nc=2.4$, with borders of the first Brillouin zone marked in orange (dashed) and the conduction electron Fermi surface in green (dash-dotted); the latter coincides with the Fermi surface of FL$^\ast$ at the mean-field level.
(c), (d) Same as (a), (b), but for higher filling $\nc=3.0$.
In both (b) and (d), the Dirac nodes of the spin-liquid component are clearly visible.
}
\label{fig:flst_bands}
\end{figure}

(iii) For large $\JK/t$ we obtain a heavy Fermi liquid (HFL), with non-zero and diagonal $\mat U$ and $\mat W$ mean-field parameters, the latter describing Kondo screening. In this phase, the topological properties of the Kitaev sector are destroyed, as discussed in more detail in Section~\ref{sec:hfl}.

(iv) Finally, there is a class of intermediate-coupling low-temperature phases which represent nematic superconductors (SC). Also here, both $\mat U$ and $\mat W$ are non-zero, but their structure is more complicated and preserves some of the properties of the Kitaev spin liquid, for details see Section~\ref{subsec:sc}.

We note that, while FL$^\ast$ is a deconfined topological phase, both FL and SC are confined phases.
Beyond mean-field theory, an additional SC$^\ast$ phase is conceivable in which the fractionalized spin-liquid component coexists with superconducting conduction electrons \cite{csbs}. Such a deconfined phase may arise via a superconducting instability of FL$^\ast$; a detailed study of this is left for future work.

\subsection{Fractionalized Fermi liquid}
\label{subsec:flst}

In the FL$^\ast$ phase, the  local moments and conduction electrons are decoupled at the mean-field level. A plot of the mean-field bandstructure along high-symmetry lines (with color-coded quasiparticle overlap) and the lowest quasiparticle band are shown in Fig.~\ref{fig:flst_bands}. As expected, FL$^\ast$ features a small Fermi volume, sharp electronic (charge $e$ and $S=1/2$) quasiparticles arising from the $c$ band, and Kitaev spin-liquid excitations carrying a $\Ztwo$ gauge charge.
Beyond mean field, the properties of the FL$^\ast$ phase can be studied in perturbation theory in $\JK$; see Section \ref{sec:pert}.

\subsection{Heavy Fermi liquid}
\label{sec:hfl}

We now turn to the heavy Fermi-liquid phase (HFL). Here we observe that the Kondo mean-field parameters can be reduced (by symmetry and gauge transformations) to the form $\mat W = a^0 \mat{1}$. The Kitaev mean fields are now identical, $u^0 = u^a = u^b$ or $\mat U = u^0 \mat{1}$, such that all $\chi^\mu$ Majoranas become dispersive; this is similar to a recent mean-field treatment of the doped Heisenberg-Kitaev model \cite{ykv12}.
Inspecting the resulting (effective) Hamiltonian $\HK \propto \iu \sum_{\langle i j \rangle} \vec \chi_i^T \vec \chi_j$ shows an invariance under $\vec\chi \rightarrow \mat{R}_G \vec\chi$ for arbitrary isospin matrices $\mat R_G$, giving rise to a manifold of equivalent mean-field solutions given by $\mat{W} \rightarrow \mat{R}_G \mat{W}$.
We therefore conclude that, given the particular structure and symmetry of the mean-field ansatz, there is no need to realize spin rotations for the Kitaev Majoranas projectively, and the quantum order of the spin liquid is destroyed.
The solution is invariant under spin rotations, as can be seen easily from Eq.~\eqref{eq:nn_spincorrel} in the case of $u^0=u^a=u^b$, and thus indicates the formation of Kondo singlets between the local moments and conduction electrons. In fact, this solution of the Majorana mean-field theory can be mapped to a more conventional slave-boson treatment as we show in Appendix \ref{sec:majToSF}.

\begin{figure}[!tb]
\includegraphics[width=\columnwidth,clip]{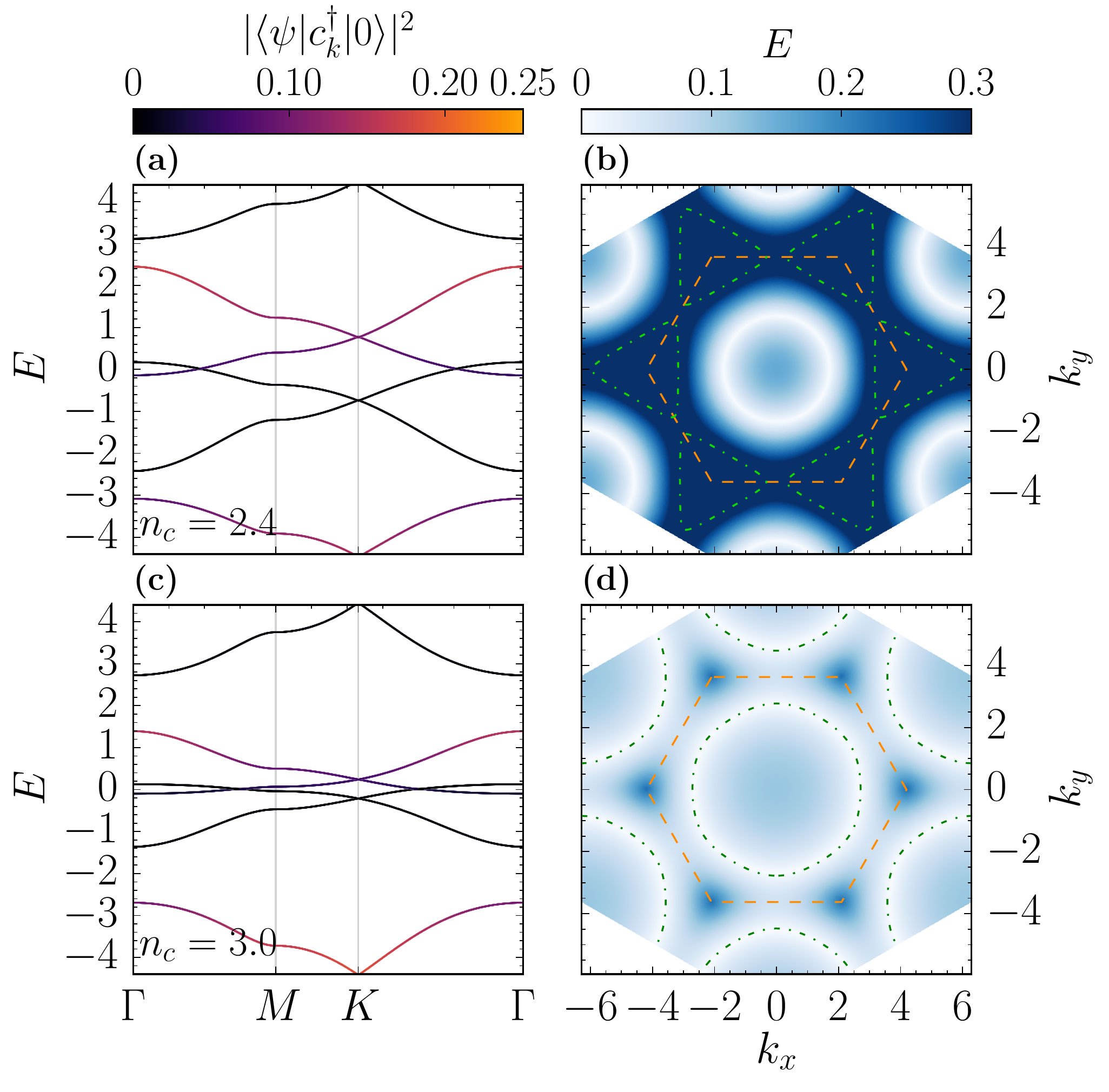}
\caption{
Mean-field band structure in the HFL phase at $\JK = 8.0$.
(a) Cut along high-symmetry lines at $\nc=2.4$ with color-coded quasiparticle weights.
(b) Energy of the lowest quasiparticle band at $\nc=2.4$ with borders of the first Brillouin zone marked in orange (dashed) and the bare conduction electron Fermi surface in green (dash-dotted).
(c), (d) Same as (a), (b), but for higher filling $n_c=3.0$.\cite{fn:tempremark}
}\label{fig:hfl_bands}
\end{figure}

The mean-field bandstructure of the HFL phase is displayed in Fig.~\ref{fig:hfl_bands}. It features a well-defined Fermi surface and rather flat bands near the Fermi level, indicating that the quasiparticles indeed have become ``heavy''.

\subsection{Superconductors}\label{subsec:sc}

The most interesting mean-field solutions are obtained at intermediate coupling $\JK$ and correspond to unconventional superconductors, with point nodes and a non-trivial structure in both momentum and spin space. For all tested values of $\nc$ and $\KK$, the solutions break the $C_3$ symmetry of combined lattice and spin rotations of the Hamiltonian: As we show below, they transform under a linear combination of two three-dimensional irreducible representations (irreps) of the group $\mathcal{S}_4$, see Appendix~\ref{sec:sym_appendix}.

\subsubsection{Mean-field parameters and symmetries}

Analyzing the symmetries of the mean-field solutions, we first note that all solutions have a $\Uone$ degeneracy, i.e., given the mean-field parameters $\mat W$, the isospin-rotated ansatz $\mat W \mat{R}_C^T$ is also a (physically inequivalent, but energy-degenerate) solution, where
\begin{equation}
  \mat{R}_C = \cos \phi \ \mat{1} + \sin \phi \ \mat{G}^3
\end{equation}
with $\phi \in [0, 2\pi)$ arbitrary.
This transformation is equivalent to taking $\vec \eta \to \mat{R}_C \vec \eta$, which corresponds to the transformation $c_\sigma \rightarrow \eu^{\iu \phi} c_\sigma$. Indeed, the phase of the anomalous expectation values $\langle c c\rangle$ changes by $2 \phi$ upon a rotation of the ansatz by $\phi$.  The $\Uone$ degeneracy in our mean-field solutions thus corresponds to the spontaneously broken $\Uone$ phase-rotation symmetry of a superconductor.
Choosing a certain phase $\phi$, we further find six distinct but energy-degenerate solutions $\ket{X_\pm}$, $\ket{Y_\pm}$ and $\ket{Z_\pm}$ which are connected by the $C_3$ operation of rotating bonds $x \to y \to z$ and spin components $S^x\to S^y \to S^z$, as described in Appendix \ref{sec:sym_appendix}, i.e. acting on the Kondo mean-field parameters as
\begin{equation}\label{eq:rot_z3}
  \mat{R}_G \mat{R}_S \mat{W}^{(X_\pm)} \mat{R}_S^T = \mat{W}^{(Y_\pm)},
\end{equation}
with the coefficients for $\mat{R}_G$ and $\mat{R}_S$ given by $a^0 = 1/2$, $a^\alpha = - 1/2$, and analogous for cyclic permutations of $(X Y Z)$. The Kitaev mean-field parameters and the kinetic energy of the electrons on the corresponding bonds will also by cyclically permuted, i.e. $u^{0,a,b}(x,y,z) \to u^{0,a,b}(y,z,x)$.
The index $+,-$ of $\ket{X_\pm}$ etc. denotes the freedom of an additional relative phase of $\pi$ between two components of the triplet vector [defined in Eq.~\eqref{eq:def_dvec} below]. Applying an appropriate $C^\ast_\alpha$ operation switches between the two solutions, e.g., $C^\ast_x \Zpm = \Zmp$, see Appendix \ref{sec:sym_appendix}.
We choose the convention that $\Zp$ corresponds to a \emph{diagonal} solution and $\Xp$ and $\Yp$ are in the orbit of the $C_3$ operation. $\Zm$ is obtained by complex conjugation of the $\arrvec d$-vector for $\Zp$, and $\Xm$, $\Ym$ lie in the orbit of $C_3$ applied to $\Zm$. Concretely, the Kondo mean-field parameters $\mat W$ for the solutions $\Zp$ of the third type are (for a suitably chosen $\Uone$ phase) diagonal and of the form
\begin{equation}
  \mat{W}^{(Z_+)} = \mathrm{diag} (a,b,b,c)
\end{equation}
where $a,b,c \in \mathbb R$. The Kitaev mean-field parameters $u^{0,a,b}(x) = u^{0,a,b}(y) \neq u^{0,a,b}(z)$ corresponding to this solution indicate a spontaneously broken spin and lattice rotation symmetry.
Alternatively, we can express any four-dimensional diagonal matrix by a linear combination of the identity matrix and products of spin and isospin matrices of equal components,
\begin{equation} \label{eq:wz_decomposed}
  \mat{W}^{(Z_+)} = b^0 \mat{1} + b^1 \mat{M}^1 \mat{G}^1 + b^2 \mat{M}^2 \mat{G}^2 + b^3 \mat{M}^3 \mat{G}^3,
\end{equation}
where the coefficients are related to the mean-field parameters by $b^0 = (a + 2 b + c)/4$, $b^1 = b^2 = (a - c)/4$ and $b^3 = (a - 2b +c) /4$.
The decoupling field $\mat W$ for the solutions of type $\Xpm$ and $\Ypm$ can be then obtained by the symmetry rotation \eqref{eq:rot_z3}, and additional reflection operations.

Hence, the superconducting solutions are not invariant under the $C_3$ rotation, i.e., they are {\em nematic}. They transform under a three-dimensional irrep of the symmetry group, see Appendix \ref{sec:sym_appendix}.
The fact that the additional multiplication with $\mat R_G$ in Eq.~\eqref{eq:rot_z3}, i.e., a gauge rotation, is required implies that the symmetry properties of the superconducting phase are influenced by the quantum order of the parent spin liquid (by which we refer to the projective realization of symmetries).
Note, however, that there is no $\mathbb Z_2$-redundancy for the $\mat U$ mean fields, as opposed to  the spin liquid or FL$^\ast$ phases. Beyond mean-field theory, the SC phase hence does not possess a $\mathbb Z_2$ gauge structure and is a topologically trivial confined phase (in contrast to a possible SC$^\ast$ phase, see Sec.~\ref{overview}).

\subsubsection{Excitation spectrum}

\begin{figure}[!tb]
\includegraphics[width=\columnwidth,clip]{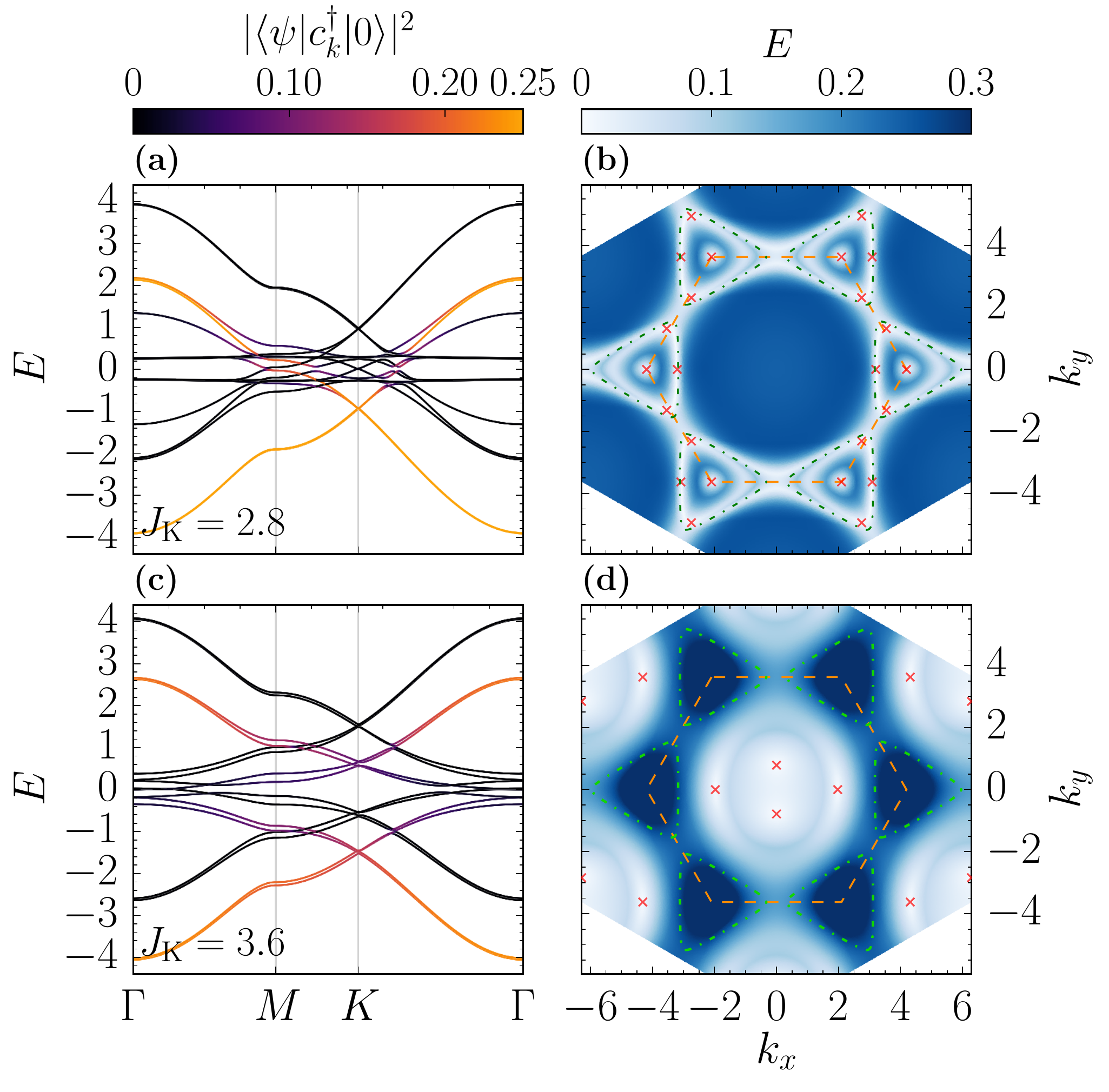}
\caption{
Mean-field band structure in the superconducting phase for $n_c = 2.4$.
(a) Cut along high-symmetry lines with color-coded quasiparticle weights close to the transition to FL$^\ast$ ($\JK=2.8$).
(b) Energy of the lowest quasiparticle band at $\JK=2.8$, with borders of Brillouin zone marked orange (dashed), nodal points marked red (crosses), and the bare conduction electron Fermi surface marked green (dash-dotted).
(c), (d) Same as (a), (b), but close to the transition the HFL ($\JK = 3.6$).
The spectrum corresponds to the solution $\ket{Z}$; the spectrum of the energy-degenerate solutions $\ket X$ and $\ket Y$ can be obtained by $\pm 2 \pi /3$ rotations around the center of the Brillouin zone.}
\label{fig:sc_1-2}
\end{figure}

\begin{figure}[!tb]
  \includegraphics[width=\columnwidth,clip]{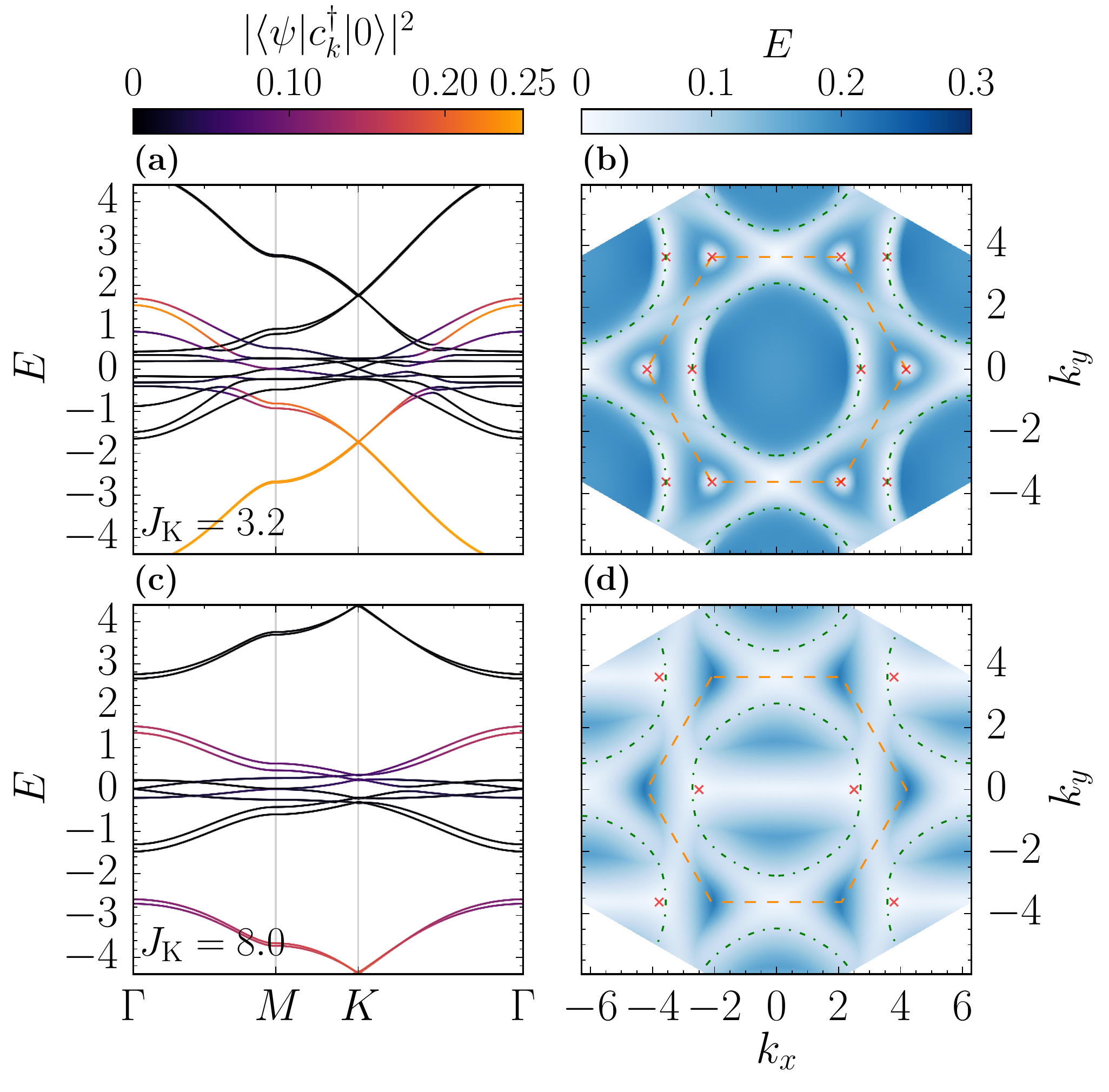}
\caption{
Same as Fig.~\ref{fig:sc_1-2}, but now for $n_c = 3.0$.
(a), (b) $\JK=3.2$. (c), (d) $\JK = 8.0$.
}
\label{fig:sc_1-5}
\end{figure}

The band structure and a plot of the lowest quasiparticle dispersion at\cite{fn:zero_temp} $T=0$ for two points in the phase diagram close to the transitions to the FL$^\ast$ and HFL phases are shown for $n_c=2.4$ and $n_c = 3.0$ in Figs.~\ref{fig:sc_1-2} and \ref{fig:sc_1-5}, respectively. The broken rotational symmetry is clearly visible from panels (b) and (d) in both figures.
For the chosen parameters, the quasiparticle energy displays multiple point nodes. Near the FL$^\ast$-SC transition the nodes are located near the original $c$-electron Fermi surface as well as very close to the $K$ points, as shown in panels a) and b) in Figs.~\ref{fig:sc_1-2} and \ref{fig:sc_1-5}, but the node count and location change continuously as a function of $\JK$. We emphasize that these nodes are accidental, and we also found regions in parameter space (e.g. around $\nc=3$, $\JK=5$) where the spectrum is fully gapped.\cite{fn:edgefoot} Technically, the appearance or disappearance of nodes corresponds to a Lifshitz transition in the superconducting state; however, most of these transitions leave only weak thermodynamic signatures, with the exception of those which are first order, see Fig.~\ref{fig:mftparams_jk} as well as Figs.~\ref{fig:pd_quant} and \ref{fig:pd_quant_fill} below.

\begin{figure}[!tb]
\includegraphics[width=\columnwidth,clip]{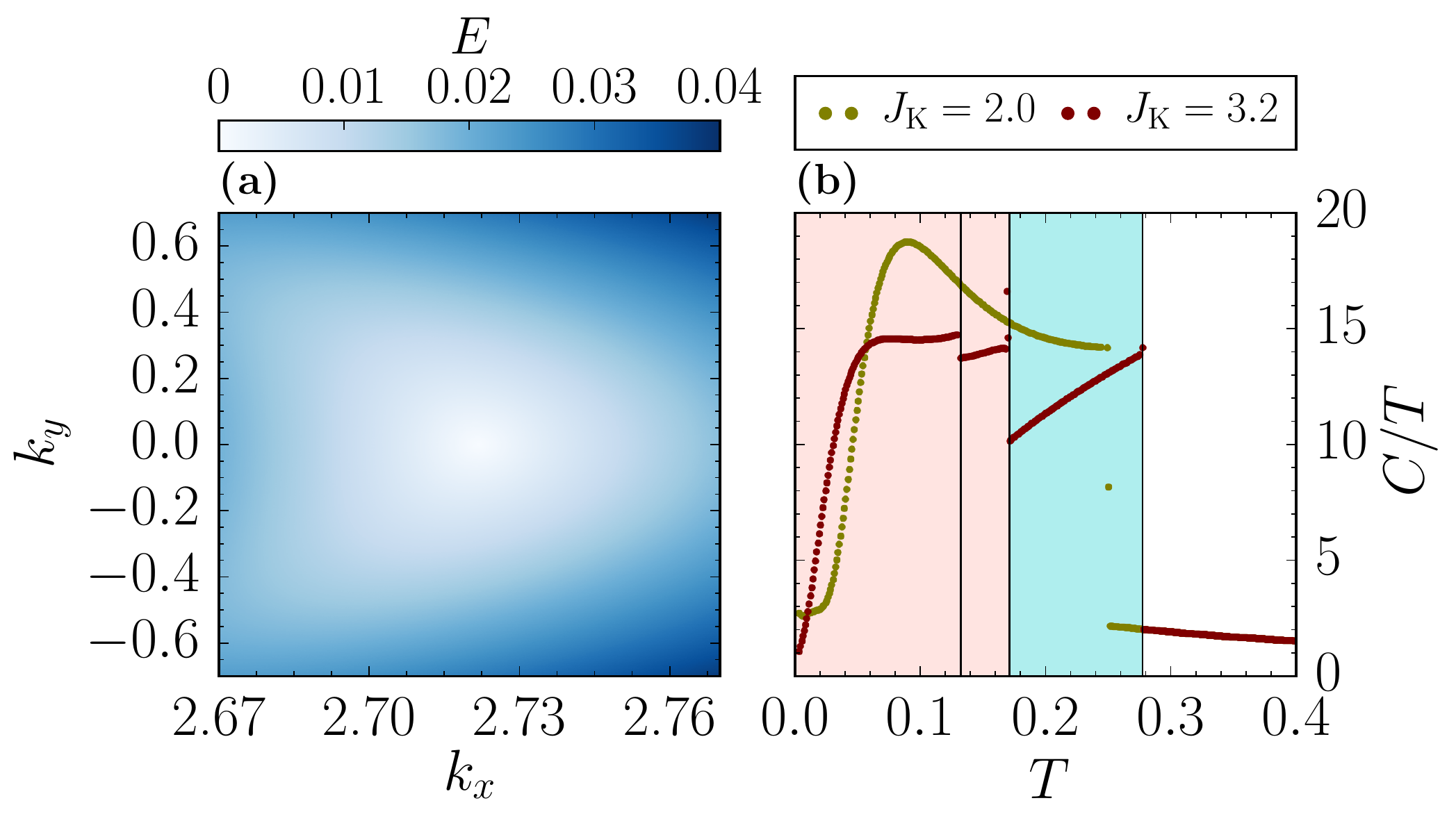}
\caption{%
(a) Momentum-space zoom into a highly anisotropic node in the superconducting phase for $\JK = 3.2$ and $\nc = 3.0$. Note that $k_x$ and $k_y$ have been rescaled to illustrate the nodal character. (b) Specific heat $C/ T$ as a function of temperature $T$ in the superconducting phase at $\nc=3.0$ and $\JK = 3.2$ [for comparision also $\JK = 2.0$ (FL$^\ast$ and decoupled regime), olive markers].\cite{fn:num_diff} The specific heat in the superconducting phase shows metallic-like behavior at intermediate $T$. Note that the indicated phase boundaries refer to the curve at $\JK = 3.2$.
}
\label{fig:aniso_node}
\end{figure}

We note that some of the nodes in the excitation spectrum have an extremely anisotropic dispersion, i.e., are characterized by two velocities which differ by 2-3 orders of magnitude.
As a result, the near-nodal energies along lines in the Brillouin zone are very small, such that the thermodynamic behavior above a small temperature scale is essentially metallic. This is illustrated in Fig.~\ref{fig:aniso_node} which shows the specific heat plotted as $C/T$ as a function of temperature.


\subsubsection{Pairing and anomalous expectation values}

For a more comprehensive symmetry analysis we compute the anomalous expectation values $\langle c_{i \sigma} c_{j \sigma'} \rangle$ of the conduction electrons on nearest-neighbor bonds and recast them into a spin-singlet component $d^0$ and a spin-triplet vector $\arrvec d$ as\cite{su91}
\begin{equation} \label{eq:def_dvec}
  \langle c_{i \sigma} c_{j \sigma'} \rangle = \left[ d^0 \iu \tau^2 + (\arrvec{\tau} \cdot \arrvec d) \iu \tau^2 \right]_{\sigma \sigma'}.
\end{equation}
We find that across all SC phases, pairing is purely triplet, i.e. $d^0 = 0$.
In a similar manner, we express the normal expectation values as $\langle c_{i \sigma}^\dagger c_{j \sigma'} \rangle = [t \tau^0 + \arrvec t \cdot \arrvec \tau]_{\sigma \sigma'}$ and observe that, depending on the solution type $\ket{\gamma} = \ket{X},\ket{Y},\ket{Z}$, the kinetic energy develops a non-zero spin component $t^\gamma$ as well as a non-zero local spin polarization in the same direction.
Exemplary results for the normal and anomalous expectation values for the solution $\Zp$ are given in Table~\ref{tab:soln_2.8}, where we use the short form $\vec d =(d^0, \arrvec d)$ and analogous for $\vec t$. We note that $\arrvec t \cdot \arrvec d = 0$ for all observed solutions.
Given the spontaneous spin polarization and that pairing is purely triplet, one might draw an analogy to the non-unitary pairing in the A1 phase of $^3 \mathrm{He}$,\cite{vowo90} however, we stress that the analogy is limited as our model is strongly spin-orbit coupled.

\begin{table}[tb]
\caption{\label{tab:soln_2.8}Components of normal and anomalous expectation values on $x,y,z$-bonds for the conduction electrons in the superconducting phase for a solution of type $\Zp$, on $x,y,z$ bonds at $T=0.03$, $\JK = 2.8$, $\KK = 4.0$, $t=1.0$ and $n_c = 1.2$. The four components of $\vec t$ occur in the decomposition $\langle c_{i \sigma}^\dagger c_{j \sigma'} \rangle = [t \tau^0 + t^\alpha \tau^\alpha]_{\sigma \sigma'}$, and $\vec d = (d^0,\arrvec d)$ denotes the singlet pairing amplitude and triplet pairing vector [cf. Eq.~\eqref{eq:def_dvec}].}
\begin{ruledtabular}
\begin{tabular}{cccccc}
  $\vec t(x)$ & $\vec t(y)$ & $\vec t(z)$ & $\vec d(x)$ & $\vec d(y)$ & $\vec d(z)$ \\
\hline
  $0.2023$ & $0.2023$ & $0.1936$ & $0$  & $0$  & $0$ \\
  $0$ & $0$ & $0$ & $-0.01936$  & $0.00365$  & $0.00771$ \\
  $0$ & $0$ & $0$ & $0.00365 \iu$  & $-0.01936 \iu$ & $0.00771 \iu$ \\
  $-0.0026$ & $-0.0026$ & $0.0123$ & $0$  & $0$  & $0$
\end{tabular}
\end{ruledtabular}
\end{table}

Having rewritten the anomalous expectation values in terms of the $\arrvec d$-vector on the $\alpha$-bonds, we can now relate the observables for the solutions $\ket X, \ket Y$ and $\ket Z$, by identifying how the transformations on the Majorana fermions $\vec \eta$ and $\vec \chi$ act on physical observables, and classify the pairing structure in terms of the symmetry group of the model.
The advantage of this approach is that the expectation values of the $c$ electrons can be regarded as physical observables, such that a discussion of the projective realization is not needed for the symmetry classification.

We find that the solutions transform in a linear combination of the two three-dimensional irreducible representations of the symmetry group of the model. For a detailed discussion, we refer the reader to Appendix \ref{sec:sym_appendix}.

\begin{figure*}[!tb]
\includegraphics[width=.9\textwidth,clip]{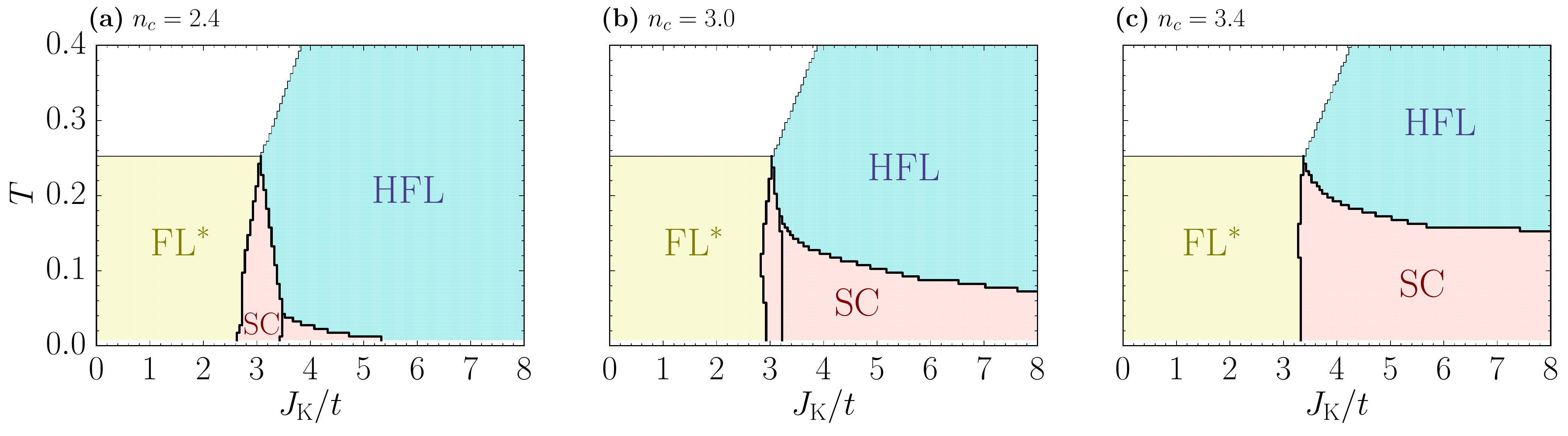}
\caption{Quantitative phase diagrams for the Kitaev-Kondo lattice, obtained from Majorana mean-field theory, as function of temperature $T$ and Kondo coupling $\JK$ for parameters $t=1$ and $\KK=4$ and different conduction-band fillings $\nc$.
(a) $\nc=2.4$,
(b) $\nc=3.0$,
(c) $\nc=3.4$.
The transitions inside the superconducting phase are accompanied by changes in the nodal structure.
Thick (thin) lines indicate first (second) order phase transitions.
}\label{fig:pd_quant}
\end{figure*}

\begin{figure}[!tb]
\includegraphics[width=0.8\columnwidth,clip]{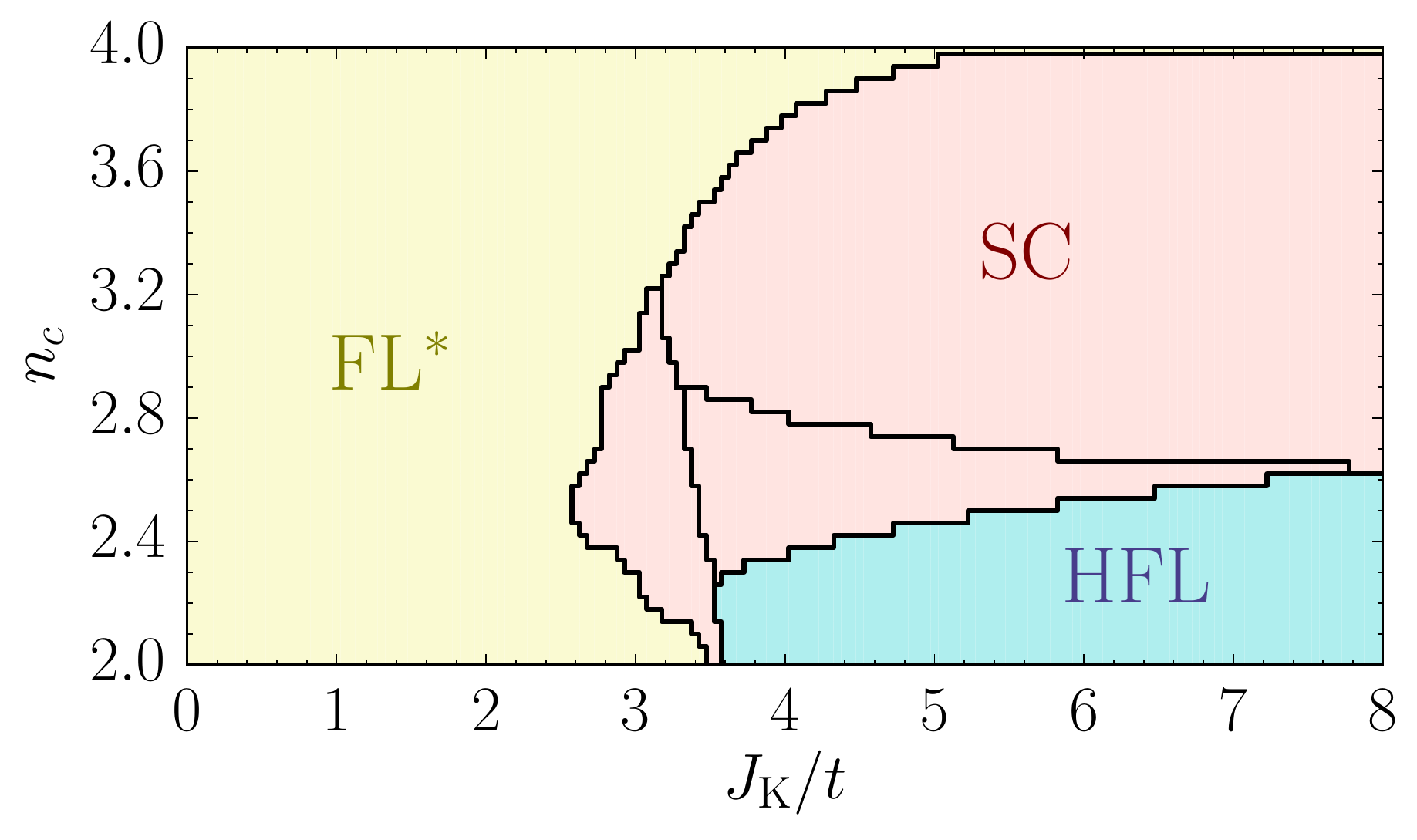}
\caption{Same as Fig.~\ref{fig:pd_quant}, but now as a function of Kondo coupling $\JK$ and conduction band filling $\nc$ at $T=0$.\cite{fn:zero_temp}
}\label{fig:pd_quant_fill}
\end{figure}

\subsubsection{Pairing glue}

One can understand the emergence of this superconducting state by integrating out the Kitaev spinons (here represented by $\vec\chi$) to obtain an effective theory for the conduction electrons.\cite{flst1} Spinons in a spin liquid have generically some finite pairing amplitude (in particular, the Kitaev model can be mapped to a $p$-wave BCS-type pairing model,\cite{cnu08}) thus inducing a finite pairing amplitude for the conduction electrons.

Interestingly, in the framework of our mean-field approach we find that mapping the Kitaev Majorana fermions to canonical fermions $f_\sigma$ results in a pairing structure for the $f_\sigma$ fermions -- even in the pure Kitaev model -- which is similar to the pairing of the conduction electrons in our SC phase (cf. Appendix \ref{subsec:spinon_pairing}). We note, however, that the $f_\sigma$ pairing itself does not correspond to superconductivity as the $f_\sigma$ do not carry charge.

This inheritance of the pairing structure can be understood as a consequence of the origin of the pairing of $c$-electrons, which is mediated by the quasiparticles in the spin liquid.
In the case at hand, the fractionalized excitations in the spin liquid are Majorana fermions,\cite{kitaev06,nasu16} such that the superconductivity in the Kitaev-Kondo lattice at intermediate $\JK$ originates from ``Majorana glue''.

\subsection{Sample phase diagrams}

Exemplary quantitative phase diagrams obtained from Majorana-fermion mean-field theory as a function of $\JK /t$ and $T$, and $\JK /t$ and $n_c$, respectively, are shown in Figs. \ref{fig:pd_quant} and \ref{fig:pd_quant_fill}. The transition from FL$^\ast$ to FL is generically masked by unconventional superconductivity, and we find that the superconducting region persists for larger $\JK$ and for higher temperatures $T$ as $\nc$ is increased.

Within the superconducting region, we often find multiple distinct solutions to the mean-field equations which have the same symmetry properties (as described above and in Appendix \ref{sec:sym_appendix}), but differ in their nodal structure and in their free energy.
By comparing the (Helmholtz) free energies of the respective solutions we determine the location of these first-order Lifshitz transition in the superconducting phase. Examples of the behavior of the mean-field parameters near these transitions are shown in Fig. \ref{fig:mftparams_jk}. We note that there are multiple additional continuous Lifshitz transitions associated with changes in the nodal structure, which we have not mapped out in detail.
We also note that the overall properties of the mean-field solutions do not appear to be affected by proximity to the Van Hove filling of the conduction band ($n_c=1.5$ and $2.5$).

The transitions surrounding the superconducting phase are observed to be first order, while the thermal transitions out of the FL$^\ast$ and HFL phases are of second-order. As discussed previously, latter transitions will become crossovers when going beyond mean-field theory \cite{flst1}.

%%%%%%%%%%%%%%%%%%%%%%%%%%%%%%%%%%%%%%%%%%%%%%%%%%%%%%%%%%%%%%%%%%%%%%%

\section{Summary}

We have introduced and studied a honeycomb Kondo-lattice model with Kitaev interactions among the local moments. We have mapped out the phase diagram using a Majorana-based mean-field theory. While large Kondo coupling yields the expected heavy Fermi liquid (HFL), small Kondo coupling leads to a fractionalized Fermi-liquid phase (FL$^\ast$) whose properties we have studied beyond mean field.

Most interestingly, the quantum confinement transition between FL$^\ast$ and FL is masked by a novel superconducting phase. It features triplet pairing driven by ``Majorana glue'', with the pairing structure inherited from the Kitaev spin liquid. It is an electron-nematic phase breaking lattice rotation symmetry, with its superconducting order parameter transforming as a linear combination of two unusual three-dimensional irreducible representations of the symmetry group, and its excitation spectrum being either gapless or displays strongly anisotropic accidental point nodes, depending on parameters.

On the experimental front, dominant Kitaev interactions have been found in a number of insulators,\cite{cha10} most prominently {\nio} \cite{sin10,chun15} and {\rucl},\cite{plumb14,sears15,banerjee17}. Hence, an experimental realization of the Kitaev-Kondo lattice appears within reach, either by adding charge carriers via doping or by engineering layered structure where, e.g., a monolayer of {\rucl} is placed on a metallic substrate.

%%%%%%%%%%%%%%%%%%%%%%%%%%%%%%%%%%%%%%%%%%%%%%%%%%%%%%%%%%%%%%%%%%%%%%%

\acknowledgments

We acknowledge instructive discussion with P. Brydon, F. Pollmann, S. Rachel, and, in particular, with C. Timm.
This research was supported by the Deutsche Forschungsgemeinschaft via Project No. A04 of SFB 1143, GRK 1621, and via the Emmy Noether program ME 4844/1 (T.M.).

%%%%%%%%%%%%%%%%%%%%%%%%%%%%%%%%%%%%%%%%%%%%%%%%%%%%%%%%%%%%%%%%%%%%%%%
%%%%%%%%%%%%%%%%%%%%%%%%%%%%%%%%%%%%%%%%%%%%%%%%%%%%%%%%%%%%%%%%%%%%%%%
%%%%%%%%%%%%%%%%%%%%%%%%%%%%%%%%%%%%%%%%%%%%%%%%%%%%%%%%%%%%%%%%%%%%%%%

\appendix

\section{Symmetries and irreducible representations}
\label{sec:sym_appendix}

\subsection{Symmetries of the Kitaev-Kondo-lattice model}

The symmetry properties of the Kitaev-Kondo lattice can be inferred from the Kitaev model, and we restrict our attention to isotropic Kitaev and Kondo couplings.
The symmetry group of the whole lattice can be generated by three operations discussed below, two of which are generalized point-group operations (which act also in the spin sector due to spin-orbit coupling), as well as a inversion symmetry.

\begin{enumerate}
  \item {The honeycomb lattice has a $C_6$ rotation symmetry at the $\Gamma$ point.
    For our purpose of analyzing the anomalous expectation values, it is sufficient to consider the symmetry group at $K$ point which reduces $C_{6} \to C_{3}$, since the triplet pairing amplitude is odd in the sublattice index.
    From $\HK$ in \eqref{h} it is evident that the $C_3$ lattice rotation operation also needs to map the spin components $S^x \to S^y \to S^z$.

    This operation can be implemented in the Majorana formalism by multiplication of the Majorana four-vectors $\vec \chi, \vec \eta$ with the $\SOfour$ spin-matrix $\mat R_S^{C_3}= (\mat 1 - \mat M^1 - \mat M^2 - \mat M^3) /2$.
    Due to the spin-gauge locking in the Kitaev model  (cf. Section \ref{subsec:proj_symm}), also a gauge transformation (given by the identical coefficients) $\mat R_G^{C_3} = (\mat 1 - \mat G^1 - \mat G^2 - \mat G^3)/2$ needs to act such that the Kitaev Majoranas transform as $\vec \chi \to \mat{R}_G^{C_3} \mat{R}_S^{C_3} \vec \chi$.}
  \item {There is a reflection symmetry $\sigma$ across an axis perpendicular to the $x$-bonds of the lattice.
    Spin-orbit coupling requires the (unitary) spin transformation $S^x \to - S^x$, $S^y \to - S^z$ and $S^z \to - S^y$ for this operation to be a symmetry.

    This spin transformation acts on the Majorana four-vectors with $\mat R_S^{\sigma} = (\mat M^2 - \mat M^3) / \sqrt{2}$ and an analogous form for $\mat R_G^\sigma$.}
  \item {There is a further symmetry operation in the Kitaev model which amounts to inversion of a single spin component, e.g. $S^x \to -S^x$. However this transformation is not unitary, one can instead consider inverting two spin components. We denote this operation by $C^\star_x : (S^x, S^y, S^z) \to (S^x, -S^y, -S^z)$.

    This operation $C^\star_\alpha$ acts on the Majorana fermions as $\vec \chi \to \mat G^\alpha \vec M^\alpha \vec \chi$ and $\vec \eta \to \mat M^\alpha \vec \eta$.}
\end{enumerate}
Applying the symmetry transformations to a given solution $\ket{\{X,Y,Z\}_\pm}$ results in a symmetry-transformed solution that is degenerate with respect to the free energy.
In particular, we find that the $\arrvec d$-vector on a given bond $\gamma$ transforms under a symmetry transformation $\hat S$ as $\hat{S}^{-1} \arrvec d(\gamma) \hat{S} = R \arrvec d(S(\gamma))$, where $R$ is a $3 \times 3$ representation matrix acting on the spin components.

\subsection{Symmetry properties of superconducting mean-field phase}

To find the irreducible representation under which our SC solutions transform, we consider the symmetry group of the model to be generated by the elements $C_3, \sigma, C^\ast_x$, as detailed above. Note that it is sufficient to determine the symmetry properties of the $\arrvec d$-vector with the respect to the symmetry group of the $K$ point, since all further symmetry properties can be deduced from requiring the anti-symmetry of the gap with respect to the sublattice index.

By arranging the $\alpha$-components of the $\arrvec d$-vector on bond $\gamma$ in a $3 \times 3$ matrix with elements
\begin{equation}
  [ d^\alpha_{\phantom {\alpha}\gamma} ] = \begin{pmatrix}
    d^x_x & d^x_y & d^x_z \\ d^y_x & d^y_y & d^y_z \\ d^z_x & d^z_y & d^z_z
  \end{pmatrix},
\end{equation}
it is easy to see that the generators fulfill the relations
\begin{subequations} \label{eq:group_rel}
  \begin{align}
    C_3 C^\ast_y &= C^\ast_z C_3 \ \text{with cyclic perm. of ($xyz$)} \\
    \sigma C^\ast_x &= C^\ast_x \sigma \quad \text{and} \quad \sigma C^\ast_y = C^\ast_z \sigma, \\
    \text{and}& \quad \sigma C_3 = C_3^2 \sigma.
  \end{align}
\end{subequations}
In particular, we find that the generators above transform the states $\ket{\{X,Y,Z\}_\pm}$ introduced in Section \ref{subsec:sc} as
\begin{subequations}
  \begin{align}
    C_3 : \{ \Xpm, \Ypm, \Zpm \} &\mapsto \{ \Ypm, \Zpm, \Ypm \} \\
    \sigma : \{ \Xpm,\Ypm,\Zpm \} &\mapsto \{ \mp \iu \Xmp, \mp \iu \Zmp, \mp \iu \Ymp \} \\
    C^\ast_x : \{\Xpm, \Ypm,\Zpm \} &\mapsto \{-\Xpm, -\Ymp, \Zmp \}.
  \end{align}
\end{subequations}
One may introduce representation matrices of the generators, acting on the $6$-dimensional representation space spanned by above states, and verify the group relations given in Eq.~\eqref{eq:group_rel} explicitly.

Proceeding, we may give the group presentation $\langle \sigma, C_3, C^\ast_x | \sigma^2 = C_3^3 = (C^\ast_x)^2= (C_3 \sigma)^2 = (C^\ast_x \sigma)^2 = (C_3 C^\ast_x)^3 = 1 \rangle$.
It is easy to see (e.g. by identifying $\sigma \to s_1$, $C_3 \sigma \to s_2$ and $\sigma C^\ast_x \to s_3$) that this group is isomorphic to the symmetric group of degree four $\mathcal{S}_4$,\cite{rose09} given by
\begin{align}
  \mathcal{S}_4 = \langle s_1, s_2, s_3 | & s_1^2 = s_2^2 = s_3^2 = 1, \nonumber \\
  & (s_1 s_2)^3 = (s_2 s_3)^3 = (s_3 s_1)^2 = 1 \rangle. \label{eq:s4_generators}
\end{align}
This group is isomorphic to the symmetry group of a cube $\mathcal{O}$ and the tetrahedral group $\mathcal{T}_d$.
The characters of the irreducible representations of $\mathcal S_4$ as well as the characters of the observed representation are given in Table~\ref{tab:irreps_s4}.
Employing the reduction formula for decomposing a reducible representation with characters $\chi(g)$ into the $j$-th irreducible representation with character $\chi^{(j)}(g)$ \cite{lang02}
\begin{equation}
  a_j = \frac{1}{|\mathcal S_4|} \sum_{g \in \mathcal{S}_4} \left[ \chi^{(j)}(g) \right]^\ast \chi(g),
\end{equation}
we find that the observed solutions to the mean-field equations transform in the linear combination of two three-dimensional representations $\Tone$ and $\Ttwo$.

One can find basis elements for the corresponding representation spaces by computing a projector $I_j$ to the $j$th irreducible representation (of dimension $d^{(j)}$) in the basis of the group elements $g$ in the reducible representation, given by \cite{lang02}
\begin{equation}
  I_j = \frac{1}{|\mathcal{S}_4|} \sum_{g \in \mathcal{S}_4} d^{(j)} \left[ \chi^{(j)}(g) \right]^\ast g.
\end{equation}
We can then give basis states [by choosing convenient (but in principle arbitrary) linear combinations of the images of basis vectors under $I_j$] that span the representation space $V_{\Tone}$ as
\begin{subequations}
  \begin{align}
  \ket{1} &= \frac{1}{2} \left(\Xp + \iu \Zp + \Xm - \iu \Zm \right) \\
  \ket{2} &= \frac{1}{2} \left(\iu \Xp + \Yp - \iu \Xm + \Ym \right) \\
  \ket{3} &= \frac{1}{2} \left(\iu \Yp + \Zp - \iu \Ym + \Zm \right).
\end{align}
\end{subequations}
Similarly we obtain the basis states for $V_{\Ttwo}$ as
\begin{subequations}
  \begin{align}
  \ket{\bar 1} &= \frac{1}{2} \left(\Xp - \iu \Zp + \Xm + \iu \Zm \right) \\
  \ket{\bar 2} &= \frac{1}{2} \left(- \iu \Xp + \Yp + \iu \Xm + \Ym \right) \\
  \ket{\bar 3} &= \frac{1}{2} \left(-\iu \Yp + \Zp + \iu \Ym + \Zm \right).
\end{align}
\end{subequations}
Note that the coefficients of $\ket{\bar 1}$ etc. are the complex conjugated coefficients of $\ket{1}$ etc.
By inspecting the action of the group elements on the above basis vectors, it can be seen that the representation matrices acting on $V_{\Tone} \oplus V_{\Ttwo}$ are now (by construction) block-diagonal.
We emphasize that our choice of basis implies that the representation matrices $D(g)$ are now real.

\begin{table}
\caption{\label{tab:irreps_s4} Character table of the symmetric group of degree four $\mathcal S_4$ defined in \eqref{eq:s4_generators}. The group can be generated by the permutations $s_i := (i,i+1)$ for $i = 1,2,3$.}
\begin{ruledtabular}
\begin{tabular}{c|ccccc}
   &$\bm{1}$ & $ 6 \, s_1$ & $ 8 \, s_1 s_2$ & $ 6 \, s_1 s_2 s_3$ & $3 \, s_1 s_3$ \\
\hline
  $\mathrm{A}_1$ & $1$ & $\phantom{-}1$ & $\phantom{-}1$  & $\phantom{-}1$ & $\phantom{-}1$ \\
  $\mathrm{A}_2$ & $1$ & $-1$ & $\phantom{-}1$  & $-1$ & $\phantom{-}1$ \\
  $\mathrm{E}_{\phantom{0}}$ & $2$ & $\phantom{-}0$ & $-1$ & $\phantom{-}0$ & $\phantom{-}2$ \\
  $\Tone$ & $3$ & $\phantom{-}1$ & $\phantom{-}0$  & $-1$ & $-1$ \\
  $\Ttwo$ & $3$ & $-1$ & $\phantom{-}0$ & $\phantom{-}1$ & $-1$\\
  \hline
  $\mathrm{T}_1 \oplus \mathrm{T}_2$ & $6$ & $\phantom{-}0$ & $\phantom{-}0$ & $\phantom{-}0$ & $-2$
\end{tabular}
\end{ruledtabular}
\end{table}

%%%%%%%%%%%%%%%%%%%%%%%%%%%%%%%%%%%%%%%%%%%%%%%%%%%%%%%%%%%%%%%%%%%%%%%

\section{Perturbation theory in $\JK$ in the FL$^\ast$ phase}
\label{append:perturbation_theory}

This appendix supplements Section~\ref{sec:pert}, discussing aspects of the perturbative treatment of the Kondo coupling $\JK$ in the fractionalized Fermi liquid phase.

As explained in the main text, the application of $\HJ$ changes the flux sector. Applied to a flux-free state, it creates two fluxes. Focussing on an effective theory within the lowest flux sector, the leading effect of $\HJ$ can thus be found in second-order perturbation theory, and corresponds to a process in which two neighboring spins communicate via the exchange of a particle-hole pair. In this process, the first electron-spin interaction creates two fluxes on the hexagons next to the link connecting these neighboring spins, which are then annihilated by the second electron-spin interaction. One can formally derive this process by integrating out the electrons. (Note that the first-order term is proportional to the expectation value of the electron spin and vanishes by time-reversal symmetry).
At second order one obtains a retarded exchange coupling of the form
\begin{align}
\mathcal{S}_{2}&=\int d\tau d\tau'\,\sum_{ ij}\sum_{\alpha,\beta}S_i^\alpha(\tau)\,\chi_{ij}^{\alpha\beta}(\tau-\tau')\,S_j^\beta(\tau').
\end{align}
with
\begin{align}
\chi_{ij}^{\alpha\beta}&(\tau-\tau')= \frac{1}{4}\sum_{\sigma\sigma'}\sum_{\bar{\sigma}\bar{\sigma}'}\,\JK^\alpha\, \JK^\beta \tau^\alpha_{\sigma\sigma'}\,\tau^\beta_{\bar{\sigma}\bar{\sigma}'}\nonumber\\
&\times\langle T_\tau\, c_{i\sigma}^\dagger(\tau)  c_{j\bar{\sigma}'}(\tau') \rangle_c\,\langle T_\tau\,c_{j\bar{\sigma}}^\dagger(\tau')  c_{i\sigma'}(\tau) \rangle_c,
\end{align}
where $\tau$ and $\tau'$ denote imaginary times. Now introducing a projector onto the flux-free sector $\Pi_0$, and dropping global energy shifts, we find the Kondo coupling to generate an exchange of the same form as the original Kitaev coupling,
\begin{align}
\Pi_0\,\mathcal{S}_{2}\,\Pi_0=\int d\tau d\tau'\,\sum_{\langle ij \rangle_\alpha}S_i^\alpha(\tau)\,\chi_{ij}^{\alpha\alpha}(\tau-\tau')\,S_j^\alpha(\tau').
\end{align}
If the Kitaev coupling is much smaller than the  electronic bandwidth, $|\KK|\ll| t|$, one can approximate $\Pi_0\,\mathcal{S}_{2}\,\Pi_0$ by its instantaneous part. This yields a correction of the order $\JK^2/t$ to the Kitaev exchange $\KK$, with a numerical prefactor that depends on the chemical potential. Since the conduction electrons hop on a honeycomb lattice, this correction can be calculated analogously to the RKKY exchange in graphene.\cite{ss_11_2}

If, on the contrary, the flux gap is not small compared to the electronic bandwidth, the electron dynamics cannot be considered faster than the spin dynamics: while the time scale for electronic hopping between neighboring sites is set by their inverse bandwidth, virtual fluctuations of the flux sector have a typical time scale of $1/\KK$. This implies that retardation effects need to be taken into account for $t\not\gg\KK$, which in turn leads to an additional suppression of the exchange coupling due to the energy of the intermediate state with two fluxes. An estimate of this additional suppression can be derived from second-order Rayleigh-Schr\"odinger perturbation theory: For an eigenstate $|n_0\rangle$ of $\mathcal{H}_0$ with $\mathcal{H}_0 \, |n_0\rangle = E_{n,0}\,|n_0\rangle$ in the flux-free sector, the second-order correction to its energy is of the form
\begin{align}
E_{n,2}&=\langle n_0|\HJ\,\frac{1}{E_{n,0}-\mathcal{H}_0}\,\HJ|n_0\rangle\nonumber\\
&=\sum_{m\neq n}\frac{|\langle n_0|\HJ\,|m_0\rangle|^2}{E_{n,0}-E_{m,0}}.
\end{align}
The energy difference of the initial state $|n_0\rangle$ and the intermediate state $|m_0\rangle$ arises from the creation of an electron-hole pair (which mediates the RKKY-type exchange), and the creation of two fluxes. Denoting the energy of a particle-hole pair with hole momentum $q$ and electron momentum $q'$ by $\epsilon(q,q')>0$, and the energy of two fluxes in the intermediate state by $\epsilon_\Phi>0$ (hence $\epsilon_\Phi\sim \KK$), we find
\begin{align}
E_{n,2}&\sim \sum_{q,q'}\frac{\JK^2}{\epsilon(q,q')+\epsilon_\Phi}\sim \frac{\JK^2}{{\rm max}\{t,\KK\}},
\end{align}
where the summation is restricted to momenta $q$ ($q'$) that are occupied (empty) in the state $|n_0\rangle$.

We conclude that the proper scaling of the second-order correction to the Kitaev coupling entering the Majorana dynamics in the lowest flux sector is given by $\JK^2/{\rm max}(t,K)$.

%%%%%%%%%%%%%%%%%%%%%%%%%%%%%%%%%%%%%%%%%%%%%%%%%%%%%%%%%%%%%%%%%%%%%%%

\section{Mean-field theory for the heavy Fermi liquid}
\label{sec:majToSF}

The purpose of this Appendix is to show that our Majorana mean-field description of the heavy Fermi liquid is equivalent to that using conventional slave-boson mean-field theory.

To this end, we consider a mean-field solution of the form $\mat W = a \mat{1}$.
This is a (for our purposes) convenient choice of ansatz, since the heavy Fermi liquid is invariant under $\mat W \to \mat R_M \mat R_G \mat W \mat R_C^T$, where $\mat R_M$ is a spin-rotation matrix (with the invariance corresponding to the spin-rotation invariance of the HFL), $\mat R_G$ an arbitrary isospin matrix (resulting from the redundancy $\vec \chi \to \mat R_G \vec \chi$) and $\mat R_C=\cos \phi \mat{1} + \sin \phi \mat{G}^3$ is a $\Uone$ symmetry transformation of the conduction electrons (which is broken in the SC phase, cf. Section \ref{subsec:sc}).

The Kondo interaction term in the mean-field approximation then reads (omitting the site index) $\HK = -3 \JK a / 4 \sum_\lambda \iu \chi^\lambda \eta^\lambda$.
Inserting the expressions for the Majorana fermions in terms of slave fermions (cf. Section~\ref{sec:mf}), we obtain
\begin{equation} \label{eq:majTosf}
  \HK = -\frac{3 \JK}{4} \iu a \left[ f_\uparrow^\dagger c_\uparrow + f_\downarrow^\dagger c_\downarrow \right] + h.c.
\end{equation}
Note that the factor $3/4$ is usually not obtained in large-$N$ treatments of the Kondo lattice since terms of the form $(1/N) c_\alpha^\dagger c_\alpha f_\beta^\dagger f_\beta$ (implicit sum over $\alpha,\beta=-N \dots N$) are not decoupled explicitly, but rather absorbed in a redefinition of the chemical potential.\cite{col83}

Since the mean-field parameter $a= \langle \iu \chi^\lambda \eta^\lambda \rangle \in \mathbb R$ for $\lambda = 0, \dots, 3$, we can rewrite the Majorana mean-field as
\begin{equation}
  a = \frac{\iu}{2} \left[ \langle f_\sigma^\dagger c_\sigma \rangle + \langle f_\sigma c_\sigma^\dagger \rangle \right]
\end{equation}
for $\sigma = \uparrow,\downarrow$.
The second term in the brackets above is the negative of the complex conjugated first term, and therefore $a$ being real (as easily seen from the Majorana representation) implies that the diagonal gauge for $\mat W$ chosen above is one such that the mean fields $\langle f_\sigma^\dagger c_\sigma \rangle$ are purely imaginary, and thus $a= \iu \langle f_\sigma^\dagger c_\sigma \rangle = - \iu \langle c^\dagger_\sigma f_\sigma \rangle$.

Hence $\HK$ in Eq.~\eqref{eq:majTosf} with $a$ expressed with canonical fermions reproduces the usual mean-field decoupling using a auxiliary-fermion/slave-boson formalism with spin-isotropic mean fields.\cite{koli88,uble92}

%%%%%%%%%%%%%%%%%%%%%%%%%%%%%%%%%%%%%%%%%%%%%%%%%%%%%%%%%%%%%%%%%%%%%%%

\section{Pairing of spinons in the Kitaev model}\label{subsec:spinon_pairing}

It is instructive to investigate the nature of the pairing of spinons in the mean-field treatment of the Kitaev model.
While the anomalous propagators and expectation values in the spin-liquid phase are unobservable, we will see that the pairing structure in the superconducting phase is rather similar.

\begin{table}
\caption{\label{tab:example_soln}Components of normal and anomalous expectation values for the slave fermions in the spin-liquid phase, on $x,y,z$ bonds with $\langle f_\sigma^\dagger f_{\sigma'} \rangle = t^\mu \tau^\mu_{\sigma \sigma'}$ and $\langle f_\sigma f_{\sigma'} \rangle = d^\mu ( \tau^\mu \iu \tau^2)_{\sigma \sigma'}$ (implicit sum over $\mu=0,\dots,3$).}
\begin{ruledtabular}
\begin{tabular}{llllll}
  $\vec t(x)$ & $\vec t(y)$ & $\vec t(z)$ & $\vec d(x)$ & $\vec d(y)$ & $\vec d(z)$ \\
\hline
  $0.0594$ & $0.0594$ & $0.0594$ & $0$  & $0$  & $0$ \\
  $0$ & $0$ & $0$ & $-0.0594$  & $0.1906$  & $0.1906$ \\
  $0$ & $0$ & $0$ & $0.1906  \iu$  & $- 0.0594 \iu$ & $0.1906 \iu$ \\
  $-0.01906$ & $-0.1906$ & $0.0594$ & $0$  & $0$  & $0$
\end{tabular}
\end{ruledtabular}
\end{table}

Rewriting $\HK$ in terms of slave fermions, by employing the inverse of the mapping detailed in Section \ref{subsec:majrep}, the Hamiltonian splits into two parts, $\HK^\uparrow$ and $\HK^\downarrow$ with
\begin{align}
    \HK^\downarrow &= \frac{\iu u^0}{2}  \sum_{\langle i j \rangle_x} \left[ f_{i \downarrow}^\dagger f_{j \downarrow} - f_{i \downarrow} f_{j \downarrow} - h.c. \right] \notag \\
    & + \frac{\iu u^0}{2}  \sum_{\langle i j \rangle_y} \left[ f_{i \downarrow}^\dagger f_{j \downarrow} + f_{i \downarrow} f_{j \downarrow} - h.c. \right] \ \text{and} \\
    \HK^\uparrow &= \frac{\iu u^a}{2}  \sum_{\langle i j \rangle} \left[ f_{i \uparrow}^\dagger f_{j \uparrow} + f_{i \uparrow} f_{j \uparrow} - h.c. \right] \notag \\
    & + \frac{\iu u^0}{2}  \sum_{\langle i j \rangle_z} \left[ f_{i \uparrow}^\dagger f_{j \uparrow} - f_{i \uparrow} f_{j \uparrow} - h.c. \right].
\end{align}
It is thus clear that there may only be spin-triplet pairing. In particular, the pairing amplitude can be specified in real space on the three inequivalent bonds $x,y,z$,
\begin{subequations}
  \begin{align}
    \Delta_{\downarrow \downarrow} &= \frac{\iu u^0}{2} (-1,1,0) \\
    \Delta_{\uparrow \uparrow} &= \frac{\iu u^a}{2} \left(1, 1, 1-\frac{u^0}{u^a} \right).
  \end{align}
\end{subequations}
Using these results to compute the $\arrvec d$-vector for the pairing of slave fermions in the spin-liquid phase yields the results displayed in Table~\ref{tab:example_soln}.
Note that the structure of these anomalous expectation values resembles the pairing amplitudes of the conduction electrons for the (diagonal) $\Zp$ solution, as shown in Table~\ref{tab:soln_2.8}.

We note that (after adopting different conventions regarding the definition of $\vec d$) these values are identical to the mean-field parameters obtained in Ref.~\onlinecite{sbk12} by requiring self-consistency for the singlet and triplet pairing channels of the slave fermions directly.

%%%%%%%%%%%%%%%%%%%%%%%%%%%%%%%%%%%%%%%%%%%%%%%%%%%%%%%%%%%%%%%%%%%%%%%

\section{Mean-field treatment of the anisotropic Kitaev model}
\label{sec:aniso_mft_kitaev}

The purpose of this Appendix is to study to what end the mean-field theory developed by You et al. \cite{ykv12} reproduces the exact solution of the Kitaev model in the anisotropic case. We consider a decoupling using the Kitaev spin representation $S^\alpha = \iu \chi^0 \chi^\alpha$ and introduce link-dependent mean fields $u^0(\gamma) = \langle \iu \chi^0_i \chi^0_j \rangle$ and $u^a(\gamma) = \langle \iu \chi^\gamma_i \chi^\gamma_j \rangle$ on $\langle i j \rangle_\gamma$-links, yielding the mean-field Hamiltonian
\begin{equation}
  \HK = \sum_{\langle i j \rangle_\gamma} \KK^\gamma \left[  \iu u^a(\gamma) \chi^0_i \chi^0_j +\iu u^0(\gamma) \chi^\gamma_i \chi^\gamma_j - u^0(\gamma) u^a(\gamma) \right].
\end{equation}
Note that here we are using Kitaev's spin representation \eqref{eq:kitaev_rep} as opposed to the main text, where we use the more general decoupling \eqref{eq:spin_maj}. The decoupling \eqref{eq:spin_maj}, with link-dependent mean-fields $u^b$, leads to a full dimerization for strongly anisotropic Kitaev coupling, yielding non-zero mean-field parameters only on one type of bond.

The mean fields can be determined by demanding self-consistency. Since the $\chi^\gamma$ Majorana fermions remain localized to their respective bond type, the value of the mean field $u^a= \mp 0.5$ is insensitive to an anisotropy of $\KK^\gamma$.
The expectation values of the matter Majorana fermions can be determined as
\begin{align}
  u^0(\gamma) = \pm \frac{1}{N} \sum_{k \in \mathrm{BZ}/2} \cos \left(\phi(\arrvec k) - \arrvec k \cdot \arrvec n_\gamma \right),
\end{align}
where we define $\phi(\arrvec k) = \arg \sum_\alpha \KK^\alpha \eu^{\iu \arrvec k \cdot \arrvec n_\alpha}$, with the reciprocal lattice vectors $\arrvec n_x$, $\arrvec n_y$ and $\arrvec n_z \equiv 0$ for notational convenience, and $N$ is the number of unit cells.

\begin{figure}[!tb]
  \includegraphics[width=1\columnwidth]{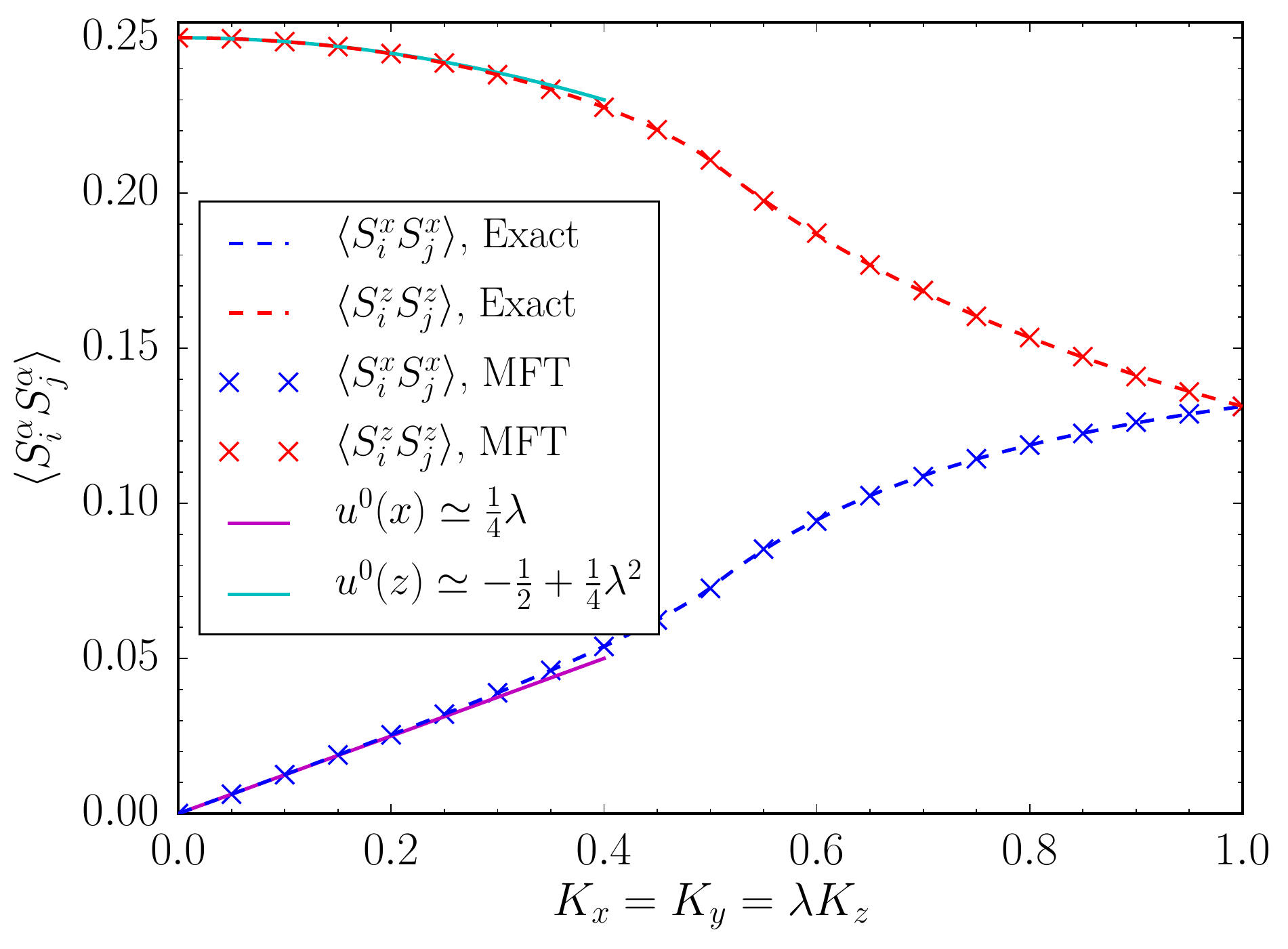}
  \caption{Static spin-correlation functions as obtained by the exact solution\cite{bms07} and by the mean-field approximation for anisotropic couplings $\KK^x = \KK^y = \lambda \KK^z$. Note that $\langle S_i^x S_j^x \rangle =\langle S_i^y S_j^y \rangle$.}\label{fig:compare_corrls}
\end{figure}

We parametrize the anisotropy as $\KK^x = \KK^y = \lambda \KK^z$ with $\lambda \leq 1$. Expanding $u^0(\gamma)$ in lowest non-trivial order of $\lambda$ yields on the respective bonds
\begin{subequations}
  \begin{align}
    u^0(x) &= u^0(y) = \frac{1}{4} \lambda + \mathcal{O}(\lambda^2) \\
    u^0(z) &= - \frac{1}{2} + \frac{1}{4} \lambda^2 + \mathcal{O}(\lambda^3).
  \end{align}
\end{subequations}
Going beyond perturbation theory, we find that the exact value\cite{bms07} of the non-vanishing static spin-correlation function $\langle S^\gamma_i S^\gamma_j \rangle$ on $\langle i j \rangle_\gamma$-links is reproduced in the mean-field treatment with
\begin{equation}
  \langle S^\gamma_i S^\gamma_j \rangle = - u^0 u^a,
\end{equation}
as can be seen in Fig.~\ref{fig:compare_corrls}.
Considering the mean-field bandstructure and the static spin correlators as shown, it is evident that the mean-field theory reproduces the exact solution.

We stress that the value of the mean-field parameter $u^0 = 0.262433$ should not be associated with the energy of the flux gap $\Delta E \simeq 0.26$, since the flux gap scales as $(\KK^x)^4/(\KK^z)^3 = \KK^z \lambda^4$ (obtained by perturbation theory in $\lambda$ on the dimer limit by Kitaev \cite{kitaev06}) for $\lambda \ll 1$, while the mean-field-parameters $u^0(x)=u^0(y) =\mathcal{O}(\lambda)$ and $u^0(z) =\mathcal{O}(1)$ in lowest order.
It is thus clear that the utility of the mean-field description is restricted to the flux-free sector, where it yields the exact matter-Majorana spectrum, whereas flat bands arising from the localized Majorana fermions do {\em not} correspond to excitations of the gauge field.

%%%%%%%%%%%%%%%%%%%%%%%%%%%%%%%%%%%%%%%%%%%%%%%%%%%%%%%%%%%%%%%%%%%%%%%

\section{Constraints and gauge transformations}\label{sec:constraints}

In this Appendix we show that the local Hilbert-space constraint $\vec \chi^T_i \mat G^3 \vec \chi_i =0$ generates the gauge operator $D_i = 4 \chi^0_i \chi^1_i \chi^2_i \chi^3_i$ as introduced by Kitaev.\cite{kitaev06}

The $D$ operator (omitting site indices) acting on the Majorana fermions can be understood as a $\Ztwo$ gauge transformation on states in Majorana basis, and is the identity on physical states $D \ket \psi = \ket \psi$.
To this end, we first consider the unitary operator $U = \exp [ \alpha \vec \chi^T \mat G^3 \vec \chi ]$.
Considering the series expansion acting on physical states $\ket \psi$, it is clear that $U$ needs to act as the identity,
\begin{equation}
  U \ket \psi = \sum_{j=0}^\infty \frac{\alpha^j}{j!} \left(\vec \chi^T \mat G^3 \vec \chi \right)^j \ket \psi = \ket \psi,
\end{equation}
since all terms with $j > 0$ annihilate $\ket \psi$. Thus only the term with $j=0$ contributes in the sum, verifying that $U$ is indeed a symmetry transformation.

We now show that the explicit resummation yields the operator $D$. We define the operators $X = \chi^0 \chi^3$ and $Y = \chi^1 \chi^2$ such that we may rewrite $\vec \chi^T \mat G^3 \vec \chi = X + Y$. In particular, we note the properties $X^2 = Y^2 = - 1/4$, $XY = D/4$ and $XD = -Y$ as well as $YD = -X$ which follow straightforwardly from the anticommutation relations of the Majorana fermions.

Using these operators, the series expansion reads
\begin{align*}
  &\eu^{\alpha \vec \chi^T \mat G^3 \vec \chi } = \eu^{\alpha (X + Y)} \\
  &= 1+  \sum_{j=1}^\infty \frac{\alpha^{2j}}{(2j)!} (X+Y)^{2j} + \sum_{j=0}^\infty \frac{\alpha^{2j+1}}{(2j+1)!} (X+Y)^{2j+1},
\end{align*}
where we already have grouped terms of even and odd powers. It is easy to prove (e.g. by induction) that $(X+Y)^{2j} = (-1)^j (X+Y)$ and $(X+Y)^{2j+1} = (-1)^j (D-1)/2$. Inserting these expressions and performing a resummation of the series, we obtain
\begin{equation} \label{eq:expToD}
  U = 1 + \vec \chi^T \mat G^3 \vec \chi \sin \alpha + \frac{1}{2} (D-1) (\cos \alpha - 1).
\end{equation}
Since the exponential on the RHS of Eq.~\eqref{eq:expToD} is the identity on physical states (as reasoned above) it follows immediately for $\alpha = \pi$ that $D \ket \psi = \ket \psi$. In fact, since $\vec \chi^T \mat G^3 \vec \chi = 0$ for physical states, the operator $D=4 \chi^0 \chi^1 \chi^2 \chi^3$ is generated along with the Hilbert-space constraint $D \ket \psi= \ket \psi$ for all values $\alpha$.

%%%%%%%%%%%%%%%%%%%%%%%%%%%%%%%%%%%%%%%%%%%%%%%%%%%%%%%%%%%%%%%%%%%%%%%
%%%%%%%%%%%%%%%%%%%%%%%%%%%%%%%%%%%%%%%%%%%%%%%%%%%%%%%%%%%%%%%%%%%%%%%
%%%%%%%%%%%%%%%%%%%%%%%%%%%%%%%%%%%%%%%%%%%%%%%%%%%%%%%%%%%%%%%%%%%%%%%


\begin{thebibliography}{99}
\bibitem{fn:maj_norm}
Note that $\HK$ is defined with spin operators $S^\alpha$, which implies that the Majorana fermions are normalized to $(\chi)^2 = 1/2$, whereas in Ref.~\onlinecite{kitaev06} the model is defined with Pauli matrices, with the Majorana fermions then normalized to unity.
\bibitem{fn:spinmap}
Another common Majorana representation of a spin,\cite{cmt94} $S^\alpha = -(\iu/2) \epsilon^{\alpha \beta \gamma} \chi^\beta \chi^\gamma$ is recovered if the sum of the ``spin'' and ``isospin'' operators from Section~\ref{subsec:majrep} is considered.
\bibitem{fn:zero_temp}
Note that all data and plots shown are obtained at $T=0.02$ (if not stated otherweise) for numerical convenience. We have checked that the values represent the $T \to 0$ limit.
\bibitem{fn:edgefoot}
For the fully gapped superconductor, we have calculated the quasiparticle spectrum for a slab geometry and did not find in-gap edge modes, indicating that the band structure is topologically trivial.
\bibitem{fn:tempremark}
The spectrum shown in panels (c,d) has been obtained at $T=0.08$.
\bibitem{fn:num_diff}
Some outlying data points are due to inaccuracies when performing the numerical differentiation at the phase transitions.
\end{thebibliography}
\end{document}